\documentclass[12pt, twoside, here]{article}
\usepackage{epsf}
\usepackage{times,colordvi,amsmath,epsfig,float,color,multicol}
\usepackage{graphics}
\usepackage{hhline}
\usepackage[large]{subfigure}
\usepackage[latin1]{inputenc}
\usepackage{rotating}
\usepackage{esint}

\newtheorem{assumption}{\sc Assumption}

\title{An Improved Temporal Formulation of Pupal Transpiration in {\em Glossina}}

\author{S. J. Childs \\ \\ {\small\em Department of Mathematics and Applied Mathematics, University of the Free State,} \\ {\small\em P.O. Box 339, Bloemfontein, 9300, South Africa.} \\ {\small\em tel: +27 51 4013386, email: simonjohnchilds@gmail.com}}

\renewcommand{\thefootnote}{\fnsymbol{footnote}}
\date{Mathematical Biosciences, 262: 214--229, 2015}

\begin{document}

\maketitle
\renewcommand{\thefootnote}{\arabic{footnote}}

\begin{abstract}
\noindent {\em The temporal aspect of a model of pupal dehydration is improved
upon. The observed dependence of pupal transpiration on time is attributed to an
alternation between two, essential modes, for which the deposition of a thin,
pupal skin inside the puparium and its subsequent demise are thought to be
responsible. For each mode of transpiration, the results of the
Bursell\nocite{Bursell1} (1958) investigation into pupal dehydration are used as
a rudimentary data set. These data are generalised to all temperatures and
humidities by invoking the property of multiplicative separability. The problem,
then, is that as the temperature varies with time, so does the metabolism and
the developmental stages to which the model data pertain, must necessarily
warp. The puparial-duration formula of Phelps and
Burrows\nocite{phelpsAndBurrows1} (1969) and Hargrove\nocite{Hargrove3} (2004)
is exploited to facilitate a mapping between the constant-temperature time
domain of the data and that of some, more general case at hand. The resulting,
Glossina morsitans model is extrapolated to other species using their relative
surface areas, their relative protected and unprotected transpiration rates and
their different fourth instar excretions (drawing, to a lesser extent, from the
data of Buxton and Lewis\nocite{BuxtonAndLewis1}, 1934). In this way the problem
of pupal dehydration is formulated as a series of integrals and the consequent
survival can be predicted. The discovery of a distinct definition for
hygrophilic species, within the formulation, prompts the investigation of the
hypothetical effect of a two-day heat wave on pupae. This leads to the
conclusion that the classification of species as hygrophilic, mesophilic and
xerophilic is largely true only in so much as their third and fourth instars are
and, possibly, the hours shortly before eclosion.} 
\end{abstract}


\section{Introduction}

Early mortality is considered to be the most significant, by far, in any model
of tsetse population dynamics (Hargrove\nocite{Hargrove1}\nocite{Hargrove3},
1990 and 2004) and the vastly different dynamics of pupal and teneral
transpiration afford each the status of a topic in its own right. While both are
crucial in deciding the viability of any tsetse population, the literature
suggests pupal dehydration to be the most challenging aspect of modelling early
mortality. The consequences of pupal dehydration are in no way limited to pupal
mortality and the prospects of eclosion alone. Transpiration continues after
eclosion up until the moment the teneral fly has its first meal. The ultimate
effect of cumulative water loss on a given cohort is therefore likely to be best
assessed in terms of the proportion of original larvae which have sufficient
reserves to achieve their first feed as tenerals (the hypothesis of
Childs\nocite{Childs6}, 2014). Combined dehydration and fat loss are thought to
culminate in significant early mortality and, while rates of teneral dehydration
are generally several times higher than those characterising the pupal stage
(Bursell\nocite{Bursell1}, 1958 and 1959), the pupal rates prevail many times
longer. Water loss during the pupal phase can therefore decide the fate of the
teneral and the significance of pupal dehydration only becomes clear in the
context of the model for teneral dehydration (Childs\nocite{Childs6}, 2014).
Pupal stage mortality is of still greater relevance in the context of control
measures, since the pupal stage is neither susceptible to targets, nor aerial
spraying. 

This research is particularly concerned with modelling the temporal dependence
of pupal transpiration rates and the variation in the
metabolic time-table to which those rates pertain. The variation of the
metabolic rate is ultimately temperature dependent, consequently, so are the
temporal domains to which each mode of transpiration applies. The formula of
Phelps and Burrows\nocite{phelpsAndBurrows1}, 1969 (modified by
Hargrove\nocite{Hargrove3}, 2004) assumes a significant role in resolving this
dependence in the absence of any more detailed information. It facilitates the
formulation of both a mapping and its derivative. This research is otherwise
based on the investigations of Bursell\nocite{Bursell1} (1958) and, to a lesser
extent, the data in another authoritative work, Buxton and
Lewis\nocite{BuxtonAndLewis1} (1934). 
One of the problems with the Bursell\nocite{Bursell1} (1958) investigation is
that many of the data are not of much use in the format in which they were
presented. Many of the data are for steady humidities at
24.7~$^\circ\mathrm{C}$, a problem that is overcome by re-interpreting
the data to be a function of calculated, accumulated water loss, or vice versa.

The transpiration-rate data for {\em Glossina morsitans} may be considered to
fall into four, essential categories: Temperature-dependent data,
humidity-dependent data, history-dependent data and time-dependent data. One
observes a certain amount of corroboration between points on the respective
curves. Some of this corroboration is demanded, for example, where the data sets
intersect, however, in other instances it comes as a pleasant surprise. The
time-dependent data are a case in point. Time-dependent transpiration would
appear to be nothing more than an alternation between two basic rates. These two
essential modes of transpiration are thought to arise as a result of the
protection afforded by the deposition of a thin, relatively impermeable, pupal
skin inside the puparium and its subsequent slow, then finally precipitous,
demise. It should, nonetheless, be emphasized that this research is neither
concerned, nor reliant on any particular biological explanation for the
different rates. Two different rates are simply observed to exist. 

The Bursell\nocite{Bursell1} (1958) data are otherwise robust in the sense that
they encapsulate a myriad of effects. Apart from the more obvious, such as
vapour pressure, the data incorporate the effects of spiracular control, the
metabolic oxidation of fat to water and any other responses of the organism to
its environment; insofar as that environment may be quantified in terms of
humidity and temperature. The data are an empirical record and the net result
of a compendium of effects, including such vagaries as the exchange of heat and
fluids with the environment. The data even resolve phenomena such as an
excretion and a historical conditioning of the puparium. They therefore
circumvent the need for any intricate, mechanistic model, involving fluxes into
and out of the organism etc. all of which would be necessary were individual
metabolic processes and reactions to be considered in isolation; a strategy
otherwise known as microsimulation. 

The main challenges to exploiting the data for the purposes of a model are in
six, very specific respects, namely, the humidity and temperature dependence of
transpiration, the historical conditioning of sensu strictu pupal transpiration,
the temporal dependence of transpiration, the variation of the metabolic
time-table with temperature, extrapolating the {\em G. morsitans}-based model to
the rest of the {\em Glossina} genus and linking total water loss to observed
pupal emergence. In the latter instance, the challenge could be described more
specifically as utilizing the dependence of survival on humidity, at
24~$^\circ\mathrm{C}$, when only the total water loss is known. A similar
problem pertains to the historical conditioning undergone prior to and at the
beginning of the sensu strictu pupal stage. The final formulation, hence the
solution to the problem, is predicated on six important assumptions. Two others
are taken for granted, in addition to those explicitly stated and explored. The
first is that the Bursell\nocite{Bursell1} (1958) investigation is
comprehensive, to the extent that it encapsulates all salient aspects of pupal
water loss. The second is that there is no transpiration at dewpoint. The
problem is then reduced to a series of integrals which can be performed on the
original, constant-temperature, time-domain of Bursell\nocite{Bursell1} (1958).
Although these integrals are extremely simple, they are both numerous and
voluminous and there are issues pertaining to differentiability and continuity.
Since a high degree of accuracy from the data, itself, is not expected and the
model is not intractably large, expedience takes precedence over taste and the
midpoint rule is the preferred integration technique. A least squares fit,
Newton's method and a half interval search are the only other numerical
techniques employed. 

The aims and broader applications of this research, in order of priority, are: 
\begin{enumerate}
\item The completion of the most challenging compartment of an early mortality model.
\item Pupal habitat assessment.
\item A better comprehension of the tsetse pupa's biology (particularly the Bursell\nocite{Bursell1}, 1958, endeavour). 
\end{enumerate}
The main causes of early mortality could be summed up as dehydration, fat loss,
predation and parasitism. Most of the experimental work needed for a model of
early stage mortality has long been complete, notwithstanding that
Bursell\nocite{Bursell2} (1959) would appear to have a small amount of data
outstanding insofar as teneral water loss' dependence on temperature is
concerned and Bursell\nocite{Bursell3} (1960) and Phelps\nocite{Phelps1} (1973)
lack a few data points pertaining to teneral fat consumption's dependence on
activity. Rogers and Randolph\nocite{RogersAndRandolph1} (1990) present a
limited amount of data linking predation and parasitism to the density at pupal
sites, an observation corroborated by \mbox{Du Toit}\nocite{DuToit} (1954). 

So far as habitat assessment is concerned, this author is by no means the first
to postulate the existence of localised and highly confined sites in which some
species larviposit. Du Toit\nocite{DuToit} (1954) concluded that such sites
existed and attributed the successful extirpation of {\em Glossina pallidipes}
from KwaZulu-Natal to focusing their efforts on the Mkhuze region. Although the
existence of the pupal sites are not specifically attributed to soil humidity,
there can be no doubt as to the motive for the
Bursell\nocite{Bursell1}\nocite{Bursell2} (1958 and 1959) investigations and
Glasgow\nocite{Glasgow1} (1963) identified river terraces as being of key
importance in the control of tsetse. Rogers and
Robinson\nocite{RogersAndRobinson} (2004) found that cold cloud duration was far
and away the most frequently occurring variable in their top five for
determining the distribution of both the {\em fusca} and {\em palpalis} groups,
using satellite imagery. Normalized difference vegetation index (NDVI) ranked
second by a significant margin in those two groups and only just beat cold cloud
duration for the {\em morsitans} group. It is not too great a stretch of the
imagination to entertain the possibility that cold cloud duration and NDVI
translate directly into soil humidity, as might elevation in the context of
river basins, vleis and low-lying, coastal areas, through which rivers typically
meander before terminating in estuaries. Rainfall was also found to be even more
relevant when it came to abundance, as opposed to distribution. 

Habitat assessment and tsetse biology are intricately entwined. That the third
and fourth instar larva should be more vulnerable to dehydration than at the
sensu strictu pupal stage is already apparent from the Bursell\nocite{Bursell1}
(1958) data. Appearances can, however, be deceptive and just how vulnerable the
organism is, is something which should not be considered in terms of individual
stages in isolation. Certainly there are surprises in store so far as the
mesophilic and xerophilic species are concerned. Making a distinction between
the hygrophilic, mesophilic and xerophilic species on the basis of their
response to a hypothetical heat wave proves to be an interesting exercise. The
implications of this research for habitat assessment only become properly
apparent in the context of the teneral model of Childs\nocite{Childs6} (2014).
It could partly explain why South Africa's sympatric, {\em Glossina
austeni}-\mbox{\em Glossina brevipalpis} population persists to this day.

\section{The Dependence of Transpiration on Temperature, Humidity and Historical Water Loss} \label{temperatureHumidityAndHistoricalWaterLoss}

The same transpiration dependences on temperature, humidity and historical water
loss are used as in Childs\nocite{Childs2} (2009). A brief summary of their
derivation follows. Bursell\nocite{Bursell1} (1958) obtained one set of water
loss data for variable temperature, in dry air, and another for variable
humidity, at 24.7 $\pm$ 3~$^\circ\mathrm{C}$. Yet a third set of data points can
be inferred by reason. One expects no transpiration at dewpoint, regardless of
the temperature. The existence of separate, temperature-dependent and
humidity-dependent data sets lends itself favourably to the assumption of
multiplicative separability. 
\begin{assumption} \label{assumption1} 
{\bf \em The transpiration rate is a multiplicatively separable function of
humidity and temperature. Put succinctly, if $\frac{dk}{dt}$ is the
transpiration rate, then there exist two functions $\phi$ and $\theta$,
dependent exclusively on humidity and temperature respectively, so that} 
\begin{eqnarray*} \label{1}
\frac{dk}{dt}(h,T) &=& \phi(h) \ \theta(T),
\end{eqnarray*} 
{\bf \em in which $h$ denotes humidity and $T$, the temperature.} 
\end{assumption}
How reasonable is this assumption? Certainly it is consistent with, and
replicates, the third, inferred set of data points entertained above. For the
`H' of data which exists across the humidity-temperature domain, one reasonably
expects rates to be bounded by the inferred, wet-end and known, dry-end data,
furthermore, to be close to monotonic. The multiplicatively separable result,
below, is consistent with the simplest, such surface. The perceived wisdom is
that the domain of interest lies mainly between 16~$^\circ\mathrm{C}$ and
32~$^\circ\mathrm{C}$, (although atmospheric temperatures hotter than that
certainly do occur in {\em G. morsitans} country). This means that the detailed,
24.7~$^\circ\mathrm{C}$ data are only being extrapolated over $\pm$
8~$^\circ\mathrm{C}$, based on the known, dry-end data and consistent with the
inferred, wet-end data. One would also expect any unusual, capricious behaviour,
or even failure in the waterproofing, to manifest itself in dry air. The
dry-air, temperature-dependent data set is, fortunately, reasonably complete and
suggestive of behaviour which is simple, smooth and monotonic (pure exponential,
in this case). It should also be pointed out that, even in the event that
circumstances were more favourable and data in the ideal format of a grid were
being interpolated, one would still ultimately be ignorant of the behaviour
between grid points, with the possibility of an unpredictable value for some,
unique combination of humidity and temperature. Thus, in the very likely event
that water loss rates are not perfectly multiplicatively separable,
multiplicative separability should not be a bad substitute. 

\begin{figure}[H]
    \begin{center}
\includegraphics[height=11cm, angle=0, clip = true]{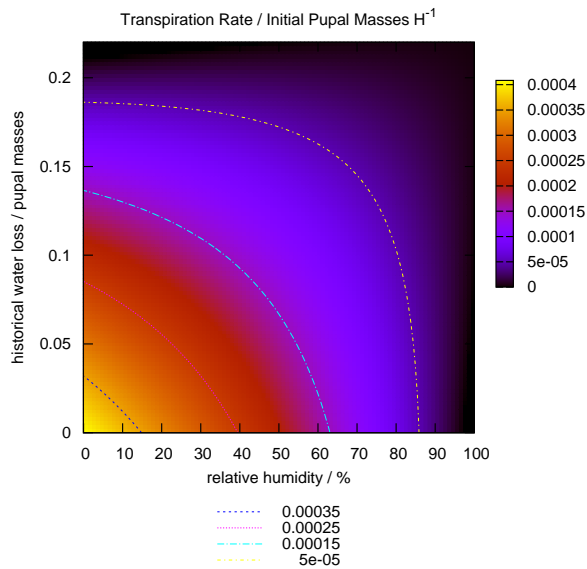}
\caption{The hourly transpiration rate as a function of humidity and water lost during the first 8/30 of the puparial duration (for {\em G. morsitans}) at 24.7~$^\circ\mathrm{C}$.} \label{historicallyConditionedWaterLoss}
   \end{center}
\end{figure}

The advantage in assuming transpiration to be a multiplicatively separable function is that the problem of its dependence on temperature and humidity is immediately reduced to a simple exercise in curve fitting, which yields 
\begin{eqnarray*} \label{2}
\phi(h) = \frac{100 - h}{100} \hspace{10mm} \mbox{and} \hspace{10mm} \theta(T) = e^{0.110268 T - 9.92201} + 0.000354783
\end{eqnarray*} 
initial-pupal-masses per hour, for the third and fourth instars
(Childs\nocite{Childs2}, 2009). Sensu strictu pupal-stage transpiration is not
quite as straightforward. Transpiration rates are determined by the temperature
and humidity which prevailed during the third and fourth instars as well as at
the beginning of the sensu strictu pupal stage itself. A second assumption
followed by minor manipulation is required in order to resolve the apparent
dependence on historical water loss.
\begin{assumption} \label{assumption2} 
{\bf \em The transpiration rate, conditioned by a given historical water loss, is the same as the transpiration rate conditioned by a historically steady humidity, at 24.7~$^\circ\mathrm{C}$, which produced an equivalent total water loss.} 
\end{assumption} 
Once this assumption has been used to quantify acclimation in terms of a historical water loss variable, the multiplicative separability assumption (Assumption \ref{assumption1}) once again facilitates a reduction of the problem to an exercise in curve fitting, which, with minor manipulation, yields
\begin{eqnarray*} 
\phi(w,h) = \frac{c_1 + c_2h + c_3w + c_4wh + c_5h^2 + c_6w^2}{\theta(24.7)} \hspace{10mm} \mbox{and} \hspace{10mm} \theta(T) = e^{0.161691 T - 12.9591}
\end{eqnarray*}
initial-pupal-masses per hour, in which $w$ is the total historical water loss
and the $c_i$ are the constants of the fit
(Childs\nocite{Childs2}\nocite{Childs2a}, 2009 and 2009a). The first formula
above describes the drought hardening reported in Bursell\nocite{Bursell1}, 1958
(in Fig. \ref{historicallyConditionedWaterLoss}, below). 
Recent, qualitative data in Fig. 4 of Terblanche and
Kleynhans\nocite{TerblancheAndKleynhans1} (2009) now suggest that acclimation
may still be possible well after the onset of the sensu strictu pupal stage.
Unfortunately, not enough is presently known to model a late acclimation
properly. There is no data suggesting how the acclimation period is composed,
what and when the onset of its effect is, or whether it is lasting and
permanent. Existing late-acclimation data were also never intended for the
quantitative purposes of a model. Problems in this regard range from obvious
inaccuracies in the presentation, sample-dependent units, unknown sample sizes,
under-weight pupae, unknown pupal history, indeterminate pupal age and the
beginnings of exponential growth in the transpiration rate at around the time of
measurement. The absorption of water vapour from the atmosphere is imagined to
be superficial (buffering by the puparium) and it is not known if a reverse
process contributes to the measured transpiration values. {\em Glossina
palpalis} transpiration rates which are lower than those for {\em G. morsitans}
are also surprising when comparing the habitats of these species, as well as
other, related data. For these and other reasons a crude late acclimation, not
commensurate with the rest of the model, was investigated and reported on
separately. Although, the omission of late acclimation renders the model
slightly deficient in detail, it seems to make little difference to the results
and it stands to reason that the later the acclimation is, the more diminished
is the benefit to the organism.

\section{The Dependence of Transpiration on Time} \label{timeDependence}

Time-dependent transpiration would appear to be little more than an alternation
between an unprotected rate and the more protected rate associated with the
sensu strictu pupal-stage. This observation is not only supported by the
temporal data (Fig. 1 of Bursell\nocite{Bursell1}, 1958), it is also
corroborated by the different temperature-dependent (Fig. 8 of
Bursell\nocite{Bursell1}, 1958) and humidity-dependent (Figs. 2, 3, 5 and 6 of
Bursell\nocite{Bursell1}, 1958) relationships which prevail prior to, then
during the sensu strictu pupal stage. In both the aforementioned data sets, a
clear distinction exists between data which pertain to the stages prior to the
commencement of the sensu strictu pupal stage and those which pertain subsequent
to its commencement. 

What is less trivial is the manner in which the Bursell\nocite{Bursell1} (1958)
time-domain must necessarily warp with fluctuations in temperature. The
assumption that parts of the puparial duration vary with temperature, in the
same way as the whole, is inevitable in the absence of any more detailed
information on the individual stages of the pupa's development. The constituent
third and fourth instars, the sensu strictu pupal stage and the pharate adult
stage are all assumed to warp in unison with the puparial duration, predicted by
the formula of Phelps and Burrows\nocite{phelpsAndBurrows1} (1969) and
\mbox{Hargrove\nocite{Hargrove3} (2004)}. A uniform dependence on an overall
metabolism for all parts of the puparial duration might be considered adequate
justification for such an assumption. Before the model can be extended to this
more general time frame, however, the temporal dependence on the original,
constant-temperature time domain of Bursell\nocite{Bursell1} (1958) first needs
to be established.

\subsection{Transpiration on Bursell's Time Domain}

Data exist for the temporal dependence of the water-loss-rate between
parturition and eclosion, at 0\% $\mathrm{r.h.}$ and 24.7 $\pm$
3~$^\circ\mathrm{C}$ (Fig. 1 of Bursell\nocite{Bursell1}, 1958). Two, essential
modes of transpiration are discernable. The rates differ by an order of
magnitude in dry air at room temperature. It is the interpretation of this
research that other, intermediate rates arise largely due to transitions between
these two, basic modes of transpiration. The two, basic rates are thought to be
a consequence of the protection, afforded by a thin, pupal skin and its
subsequent slow, then finally precipitous, demise. It should nonetheless be
emphasized that this research is neither concerned nor dependent on any
particular biological mechanism. Two different rates, with transitions between
them, are simply observed to exist.

The elevated rate, that which occurs in the absence of any waterproofing, is
\begin{eqnarray}
\label{102}
{\frac{dk}{dt}}_{\mbox{\scriptsize unprotected}}(h,T) &=& 24 \ \frac{100 - h}{100} \ ( e^{0.110268 T - 9.92201} + 0.000354783 )
\end{eqnarray}  
initial-pupal-masses per day in which $h$ denotes humidity and $T$, the
temperature (Childs\nocite{Childs2}, 2009). This rate prevails from parturition
and persists for the duration of the third and fourth instars, only to reappear
again, shortly before eclosion. The vagaries of the time-dependence during the
third and fourth instars were ignored, for want of better data, and only the
subsequent stages were modelled based on the Bursell\nocite{Bursell1} (1958)
Fig. 1 data. Transpiration plummets at the end of the fourth instar as a result
of the new-found protection afforded by the deposition of the thin, pupal skin
inside the puparium. The rate
which ultimately serves as the basis to the sensu strictu pupal stage was
determined to be 
\begin{eqnarray*} \label{103}
{\frac{dk}{dt}}_{\mbox{\scriptsize protected}}(w,h,T) &=& 24 \ (c_1 + c_2h + c_3w + c_4wh + c_5h^2 + c_6w^2) \ e^{0.161691 (T - 24.7)}
\end{eqnarray*} 
initial-pupal-masses per day, in which $w$ is the total historical water loss
and the $c_i$ are the constants of the fit (Childs\nocite{Childs2}, 2009). In
the brief period before acclimation is concluded, the same, linear
humidity-dependence as for the third and fourth instars, $\frac{100 - h}{100}$,
is temporarily assumed and used in conjunction with the protected-stage
temperature-dependence, $e^{0.161691 T - 12.9591}$. Institution of the protected
stage rate is followed by a long, slow and slight rise in transpiration rates,
from the initial minimum attained at the start of the sensu strictu pupal phase.
One possible cause is that the puparial exuviae lose some of their competence as
they age, developing minute cracks etc. as the pupa develops inside. The
transpiration rate finally makes a spectacular return to its initial levels, an
event which coincides with the cracking of the pupal case, shortly before
eclosion.

\begin{figure}[H]
    \begin{center}
\includegraphics[height=12cm, angle=-90, clip = true]{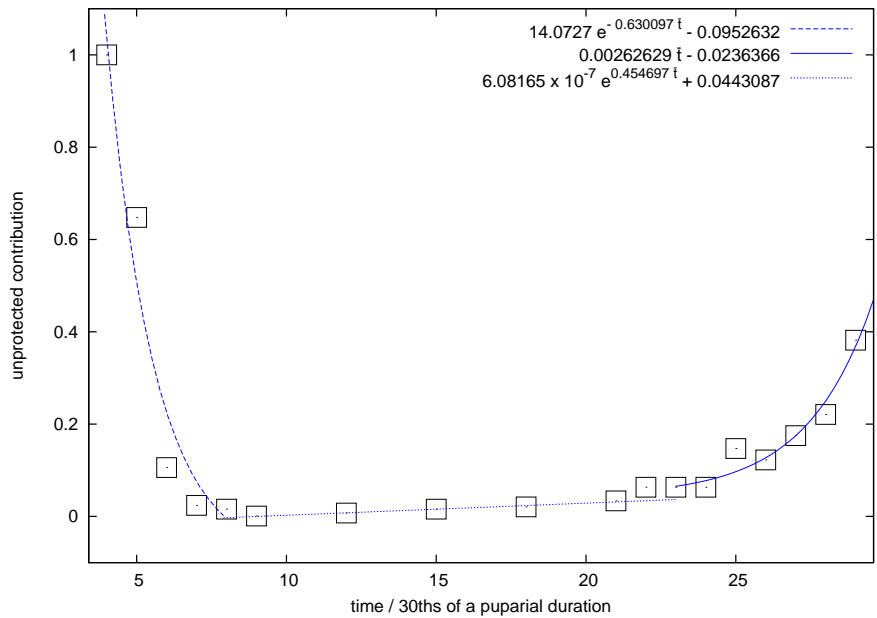}
\caption{The functions used in the successive weighted averages are the degree
to which the organism is unprotected by the thin pupal skin inside the puparium
or the proportion Eq. (\ref{102}) contributes to the transpiration rate.}
\label{mixOfRatesWithTime}
   \end{center}
\end{figure}

Three, successive, weighted averages were used to model the mix of the two
rates. The mix of the two rates for the days following the fourth instar was
determined by taking the unprotected rate to be 0.001102571 initial pupal masses
per hour and the initial pupal rate to be 0.000135555 $\mathrm{h}^{-1}$, in dry
air at room temperature. Exponential decay was deemed to produce a good model of
the transition down to the protected rate (between day four and day eight, at
room temperature), the unprotected rate then creeping in, linearly, over time
(between day eight and day 21, at room temperature). Exponential growth was used
to model the transition back to the unprotected rate (between day 21 and day 29,
at room temperature). The functions in Fig. \ref{mixOfRatesWithTime} were
consecutively used to model the relative mix of the two rates with the following
results, in which $\bar t$ denotes time in the constant temperature time frame. 
\subsubsection*{The Period ${\bar t} = ( 4, 8 ]$}
During this stage there is an adjustment from the unprotected rate down to the protected rate dictated by exponential decay, so that the transpiration rate was deemed to be
\begin{eqnarray} \label{period4to8}
{\frac{dk}{dt}} &=& \left[ 1 - \left( 14.0727 \ e^{- 0.630097 {\bar t}} \ - \ 0.0952632 \right) \right] {\frac{dk}{dt}}_{\mbox{\scriptsize protected}} \nonumber \\ 
&& + \ \left( 14.0727 \ e^{- 0.630097 {\bar t}} \ - \ 0.0952632 \right) {\frac{dk}{dt}}_{\mbox{\scriptsize unprotected}}.
\end{eqnarray}
\subsubsection*{The Period ${\bar t} = ( 8, 21 ]$}
Transpiration during this phase is predominantly at the pupal-stage rate. That is,
\begin{eqnarray} \label{period8to21}
{\frac{dk}{dt}} &=& \left[1 - \left( 0.00262629 \ {\bar t} - 0.0236366 \right) \right] {\frac{dk}{dt}}_{\mbox{\scriptsize protected}} \nonumber \\ 
&& + \ ( 0.00262629 \ {\bar t} - 0.0236366 ) {\frac{dk}{dt}}_{\mbox{\scriptsize unprotected}}. 
\end{eqnarray}
\subsubsection*{The Period ${\bar t} = ( 21, 29 ]$}
Transpiration then begins its return to the unprotected rate during the pharate adult phase. That is,
\begin{eqnarray} \label{period21to29}
{\frac{dk}{dt}} &=& \left[ 1 - \left( 6.08165 \times 10^{-7} \ e^{0.454697 {\bar t}} + 0.0443087 \right) \right] {\frac{dk}{dt}}_{\mbox{\scriptsize protected}} \nonumber \\ 
&& + \left( 6.08165 \times 10^{-7} \ e^{0.454697 {\bar t}} + 0.0443087 \right) {\frac{dk}{dt}}_{\mbox{\scriptsize unprotected}}.
\end{eqnarray} 
All three rate formulae, together with Eq. \ref{102}, constitute a sequence of
first order, ordinary differential equations which one expects to be Lipshitz
continuous over each of the stages identified (likely, even a contraction over
the greater portion of the time domain). They are the water loss rates which prevail for the constituent intervals identified on the constant-temperature time domain of Bursell\nocite{Bursell1} (1958).

\subsection{Mapping Bursell's Time Domain into Real Time}

The circumstances one is confronted with in reality are usually not those for
which the temperature is constant, let alone a constant 24.7~$^\circ\mathrm{C}$.
What if the temperature varies? The problem is that as the temperature
varies, so the metabolic process of the organism is either accelerated, or
deaccelerated. As the temperature varies, so the duration of the various
developmental stages, established on the constant-temperature time interval,
warp; stretching at low temperature and shrinking at high temperature. 

What the Bursell\nocite{Bursell1} (1958) temporal data have so far been used to
establish is the metabolic time-table for transpiration as a function of the
constant-temperature time-frame or, more succinctly, a function
$\frac{dk}{dt}(t({\bar t}),w,h,T)$ dependent on $\bar t \in [0, 30]$. This
time-frame is rooted in a particular metabolic rate, namely that associated with
a constant temperature of 24.7~$^\circ\mathrm{C}$, whereas what is of interest
is the total water loss associated with more general temperature conditions. If
time-dependent transpiration rates really can be simplistically regarded as
being dependent on the stage of pupal development and the historical
conditioning of the puparium alone (ignoring the ultimate, underlying dependence
of humidity and temperature on time for the present), then the solution to the
problem lies in a mere change in the variable of integration; should this be
possible. That is,
\begin{eqnarray*} \label{118}
\int_0^{\tau_0} \frac{dk}{dt}\left(t,w,h,T\right) dt &=& \int_0^{30} \frac{dk}{dt}\left( t({\bar t}),w,h,T \right) \frac{dt}{d {\bar t}} \ d{\bar t}, \nonumber \\ 
\end{eqnarray*}
in which $\tau_0$ is the puparial duration for the temperature history in
question. Only the lack of a knowledge of one function prevents the integral on
the right from being evaluated. Only the lack of knowledge of $t({\bar t})$
prevents the performance of the integration in the ${\bar t}$ time-frame. Since
its derivative is obviously required and since one's only knowledge of
temperature and humidity data can be expected to be in terms of the actual time,
$t$, a one-to-one, invertible mapping from the Bursell\nocite{Bursell1} (1958),
time domain, into the pertinent time-frame of some, more general case at hand,
must be formulated. By knowing only the function $t({\bar t})$, or its
derivative, $\frac{dt}{d {\bar t}}$, either one can be deduced from the other
and the correct $T(t({\bar t}))$ and $h( t({\bar t}))$ data be retrieved. 

The formula for the puparial duration is the only, potentially exploitable
information in this regard and could be the key to what is sought. The puparial
duration in days, $\tau$, has been found to vary according to the formula
\begin{eqnarray} \label{13}
\tau &=& \frac{ 1 + e^{a + bT} }{\kappa},
\end{eqnarray} 
in which $T$ is temperature (Phelps and Burrows\nocite{phelpsAndBurrows1},
1969). For females, $\kappa = 0.057 \pm 0.001$, $a = 5.5 \pm 0.2$ and $b = -0.25
\pm 0.01$ (Hargrove\nocite{Hargrove3}, 2004). For males, $\kappa = 0.053 \pm
0.001$, $a = 5.3 \pm 0.2$ and $b = -0.24 \pm 0.01$ (Hargrove\nocite{Hargrove3},
2004). The puparial durations of all species, with the exception of {\em G.
brevipalpis}, are thought to lie within 10\% of the value predicted by this
formula (Parker\nocite{Parker1}, 2008). {\em G. brevipalpis} takes a little
longer. To expect parts of  the pupa's development to be affected by temperature
in the same way as the whole, that there are no `bottle-necks' in the pupa's
development, seems a reasonable assumption in the absence of any information to
the contrary. 
\begin{assumption} \label{assumption6} 
{\bf \em The duration of any fraction of the puparial duration varies with temperature as the whole.} 
\end{assumption} 
That the duration of all the stages in the pupa's development are determined by
the same set of endothermic reactions, known collectively as the metabolism,
could be considered justification for such an assumption. At constant
temperature, a claim that the increments are related according to the formula 
\begin{eqnarray*}
d{\bar t} &=& \frac{30}{\tau(T)} \ dt
\end{eqnarray*}
would then certainly be correct. What about real-life scenarios in which the
temperature varies? The relationship between the increments is readily extended
to variable-temperature scenarios to yield
\begin{eqnarray} \label{jacobian}
\frac{dt}{d{\bar t}} &=& \frac{\tau(T(t))}{30}.
\end{eqnarray}
Once $\frac{dt}{d{\bar t}}$ has been formulated, the actual
time, $t$, corresponding to a point, $\bar t$, within the interval of
integration can be recovered from 
\begin{eqnarray} \label{integrateDerivative}
{\bar t}(t) &=& \int_{0}^{t} \frac{d{\bar t}}{dt}(t') dt' \nonumber \\
&=& \int_{0}^{t} \frac{1}{\frac{dt}{d{\bar t}}(t')} dt',
\end{eqnarray} 
in which $t'$ is a dummy variable of integration. It is in this way that a
mapping between the respective time domains is formulated to facilitate
integration on the constant-temperature time domain. It is in this way that
pertinent temperature and humidity input-data can be retrieved to facilitate
integration on the constant-temperature time domain. Should there be no
interpolation between discrete, daily temperature data, the integral in Eq.
\ref{integrateDerivative} naturally becomes a summation (usually terminating
with a fraction of a day's contribution). An exposition of this is provided in
\mbox{Section \ref{discreteTimeAndJacobian}}. 


\section{Adapting the {\em G. morsitans} Model to Other Species} \label{otherSpecies}

Transpiration-related rates for a number of species were measured at both the
unprotected and protected stages, by Buxton and Lewis\nocite{BuxtonAndLewis1}
(1934) and Bursell\nocite{Bursell1} (1958). The rates are quoted in units of
\mbox{$\mathrm{mg} \ \mathrm{h}^{-1} \mathrm{cm}^{-2} (\mathrm{mm \ Hg})^{-1}$}.
Surface-area data for pupae in the wild are likewise available and all are
tabulated in Table 6 of Bursell\nocite{Bursell1} (1958). 

In terms of fluxes, if $\rho$ denotes the density of a fluid, $\bf v$ denotes
the net velocity of that fluid through the surface of the organism, $\bf n$
denotes the normal to that surface, $P$ is a pressure and $ds$ is an element of
area, then the quantities supplied in Table 6 of Bursell\nocite{Bursell1} (1958)
are
\begin{eqnarray*}
p = \frac{\rho \oiint {\bf v} \cdot {\bf n} \ ds}{P \oiint ds} \hspace{10mm} \mbox{and} \hspace{10mm} s_{\mbox{\scriptsize species}} = \oiint ds,
\end{eqnarray*} 
whereas 
\begin{eqnarray*}
\frac{dk}{dt} = \rho \oiint {\bf v} \cdot {\bf n} \ ds = p \times s_{\mbox{\scriptsize species}} \times P,
\end{eqnarray*} 
is the transpiration rate sought. One can alternatively regard transpiration
through the integuments of each species to be governed simplistically by
something like Darcy's law, since permeability is pressure dependent. The
transpiration rate per fly is again the relevant rate multiplied by the surface
area and a pressure. 

Using either argument, a species conversion factor for unprotected-stage rates can be defined as
\begin{eqnarray} \label{11}
\delta_{\mbox{\scriptsize unprotected}} &\equiv& \frac{ p_{\mbox{\scriptsize unprotected}} }{ p_{\mbox{\scriptsize morsitans unprotected}} } \times \frac{ s_{\mbox{\scriptsize species}} }{ s_{\mbox{\scriptsize morsitans}} }, \nonumber \end{eqnarray}
in which $p_{\mbox{\scriptsize morsitans unprotected}}$ and
$p_{\mbox{\scriptsize unprotected}}$ are the unprotected-stage rates for {\em G.
morsitans} and the species in question, respectively, and $s_{\mbox{\scriptsize
morsitans}}$ and $s_{\mbox{\scriptsize species}}$ are the surface areas of {\em
G. morsitans} and the species in question, respectively. 
\begin{table}[H]
    \begin{center}
\begin{tabular}{l l|c c c}  
&  &  &  & \\
Group & Species & $\delta_{\mbox{\scriptsize unprotected}}$ & $\delta_{\mbox{\scriptsize protected}}$ (for minima) & $\delta_{\mbox{\scriptsize protected}}$ (for maxima) \\ 
&  &  &  & \\ \hline 
&  &  &  & \\
{\em morsitans} & {\em austeni} \ & 1.60 & 0.712 & 0.722 \\ 
 \ & {\em morsitans} & 1 & 1 & 1 \\ 
 \ & {\em pallidipes} & 1.50 & 1.24 & 1.31 \\ 
 \ & {\em submorsitans} & 1.59 & 0.949 & -\\ 
 \ & {\em swynnertoni} & 0.830 & 0.869 & 0.892 \\ 
&  &  &  & \\
{\em palpalis} & {\em palpalis} & 2.54 & 1.41 & 1.36 \\ 
 \ & {\em tachinoides} & 0.818 & 0.743 & - \\ 
&  &  &  & \\
{\em fusca} & {\em brevipalpis} & 10.3 & 4.57 & 3.06 \\ 
 \ & {\em fuscipleuris} & 8.84 & 4.45 & 3.16 \\ 
 \ & {\em longipennis} & 3.62 & 2.45 & 2.30 \\ 
\end{tabular}
\caption{Species conversion factors for the model calculated from data, ultimately sourced from Buxton and Lewis\nocite{BuxtonAndLewis1} (1934), presented in Bursell\nocite{Bursell1} (1958).
} \label{modelConversionFactors}
    \end{center}
\end{table}
A dimensionless, species conversion factor for sensu strictu pupal-stage rates
can be defined, similarly, as
\begin{eqnarray} \label{12}
\delta_{\mbox{\scriptsize protected}} &\equiv& \frac{ p_{\mbox{\scriptsize protected}} }{ p_{\mbox{\scriptsize morsitans protected}} } \times \frac{ s_{\mbox{\scriptsize species}} }{ s_{\mbox{\scriptsize morsitans}} }, \nonumber \end{eqnarray} 
in which $p_{\mbox{\scriptsize morsitans protected}}$ and $p_{\mbox{\scriptsize
protected}}$ are protected-stage transpiration rates for {\em G. morsitans} and
the species in question, respectively. The same surface area is used for both
the puparium and the third instar larva, the justification being that the
puparial exuviae render the puparium marginally bigger while the larval surface
is not as regular. Actual values of $\delta_{\mbox{\scriptsize unprotected}}$
and $\delta_{\mbox{\scriptsize protected}}$, for ten different species, are
tabulated in Table \ref{modelConversionFactors}. 

Notice that the ratio of the unprotected to protected conversion factors in
Table \ref{modelConversionFactors} is curiously close to two for hygrophilic
species, whereas it is around unity for both the mesophilic and xerophilic
categories. (It could suggest these categories have a layer of some, or other,
protection which is twice as thick.) Both the exceptions to this rule, {\em
Glossina submorsitans} and {\em Glossina longipennis}, occur in Sudan as well as
Ethiopia and the aforementioned ratio is around 1.5 in both. Notice, also, that
conversion factors alternatively deduced from the transpirational maxima, then
minima, during the sensu strictu pupal stage, are remarkably similar for a given
species (for all except {\em G. brevipalpis} and {\em Glossina fuscipleuris}).
This is very encouraging and immediately suggestive of a similar slope in the
time-dependence of the transpiration rate across all species. The suggestion is
that the ratios in Table \ref{modelConversionFactors} are fixed over time, {\em
G. brevipalpis} and {\em G. fuscipleuris} being the possible exceptions (a
lower, maxima-based value in the case of {\em G. brevipalpis} might be due to
the slightly longer puparial duration). The conversion factors alternatively
deduced from the transpirational maxima, then minima suggest no behavioural
differences. With this observation comes the dawning realization that the
strategies and metabolic time-table, adopted by the different species could all
be very similar and that these conversion factors might be all that is necessary
to enable both the unprotected and protected transpiration rates for another
species to be calculated from {\em G. morsitans} values. Such speculation is
further reinforced by the known, third-instar transpiration rates for both {\em
Glossina brevipalpis} and {\em Glossina palpalis}, although very little data are
available for other species. Conversion of third-instar, {\em G.
morsitans}-model, transpiration rates to {\em G. brevipalpis} and {\em G.
palpalis} values, on this basis yields errors of 6\% and 10\% respectively
(Childs\nocite{Childs2}, 2009). Finally, qualitative data, recently brought to
light in Terblanche and Kleynhans\nocite{TerblancheAndKleynhans1}  (2009),
suggest that the transpiration rates for a number of species are indeed
approximate multiples of the {\em G. morsitans} rate, during the sensu strictu
pupal stage. Such observations present a strong argument for asserting that the
only differences between the species, so far as water loss is concerned, are
permeability, surface area and a fourth-instar excretion. 
\begin{assumption} \label{assumption3} 
{\bf \em The water management strategies of the majority of tsetse fly species differ only with respect to relative pupal surface area, relative unprotected and protected transpiration rates and the different amounts excreted during the fourth instar.}
\end{assumption} 
On the face of it, Assumption \ref{assumption3} is certainly the most tenuous.
How valid is it? Does such a simplistic approach work? No dependence with
pronounced variation in temperature and humidity is indicated, since the
pressures cancel when using the original data on which \mbox{Table
\ref{modelConversionFactors}} is based. It is, nonetheless, of considerable
comfort that the conversion of the {\em G. morsitans} model to other species
involves relative rates. Assuming continuity, one should at least be able to
have confidence in results that are within a few degrees of room temperature.
The perceived wisdom is that the region of interest lies mainly between
16~$^\circ\mathrm{C}$ and 32~$^\circ\mathrm{C}$. 

\section{The Resulting Model for Pupal Water Loss}

Collecting together all prior derivation and assumptions gives rise to a series
of governing equations. Their application to any, specific temperature and
humidity data set is then facilitated by a mapping of the time domain and its
derivative alone. 

\subsection{The Governing Equations}

The total water loss of the organism is formulated as a series of five
integrals, based on a decomposition of the temporal domain and an excretion.
Note that water losses are in units of {\em G. morsitans}, initial-pupal-masses
(31 $\mathrm{mg}$). 

\subsubsection*{The Period ${\bar t} = \left[ 0, 4 \right]$}

The water loss rate for the greater part of the third and fourth instars is at the unprotected rate. Generalising Eq. \ref{102} to all species and integrating leads to
\begin{eqnarray} \label{14}
k\mid_0^4 &=& \int_0^4 24 \left( e^{0.110268 T - 9.92201}  + 0.000354783 \right) \frac{100 - h}{100} \ \frac{ p_{\mbox{\scriptsize unprotected}} }{ p_{\mbox{\scriptsize morsitans unprotected}} } \frac{ s_{\mbox{\scriptsize species}} }{ s_{\mbox{\scriptsize morsitans}} } \ \frac{dt}{d{\bar t}} \ d{\bar t}. \nonumber
\end{eqnarray}

\subsubsection*{The Period ${\bar t} = ( 4, 8 ]$}

During this period there is an adjustment from the unprotected rate down to the protected rate dictated by the exponential decay in Eq. \ref{period4to8}. Generalising to all species and integrating leads to the expression
\begin{eqnarray} \label{15}
k\mid_4^8 &=& \int_4^8 24 \left[ \left[ 1 - \left( 14.0727 \ e^{- 0.630097 {\bar t}} \ - \ 0.0952632 \right) \right] e^{0.161691 T - 12.9591} \frac{ p_{\mbox{\scriptsize protected}} }{ p_{\mbox{\scriptsize morsitans protected}} } \right. \nonumber \\ 
&& + \ \left( 14.0727 \ e^{- 0.630097 {\bar t}} \ - \ 0.0952632 \right) \left( e^{0.110268 T - 9.92201} + 0.000354783 \right) \nonumber \\ 
&& \left. \frac{ p_{\mbox{\scriptsize unprotected}} }{ p_{\mbox{\scriptsize morsitans unprotected}} } \right] \frac{100 - h}{100} \frac{ s_{\mbox{\scriptsize species}} }{ s_{\mbox{\scriptsize morsitans}} } \ \frac{dt}{d{\bar t}} \ d{\bar t}, \nonumber
\end{eqnarray}
the overall dependence on humidity being the one which exists prior to the dependence on historical water loss.

\subsubsection*{Excretion} 
If water loss is sufficiently low during the first ${\bar t} = 8$ days of the puparial duration, cognizance must be taken of excretion. In such scenarios the formula
\begin{eqnarray*} 
k\mid_0^4 + k\mid_4^8 &=& 0.0750 + \frac{h_{\mbox{\scriptsize 3rd instar}} }{ 100 } ( 0.0885 - 0.0750 ) 
\end{eqnarray*} 
was implemented for members of the {\em morsitans} group. Values of 0.0585 and
0.06 of the pupal mass were used for {\em G. brevipalpis} and {\em G. palpalis},
respectively, for want of any greater wisdom. The only fourth instar excretion
data are those for {\em G. morsitans}, {\em G. palpalis} and {\em G.
brevipalpis}. The {\em G. morsitans} rate was proportionally adjusted for other
members of the {\em morsitans} group.
 
\subsubsection*{The Period ${\bar t} = ( 8, 21 ]$}

Transpiration during this phase is predominantly at the protected rate. There is also deemed to be a small component of loss at unprotected rates, which increases linearly over time, according to Eq. \ref{period8to21}. Generalising to all species and integrating leads to
\begin{eqnarray} \label{17}
k\mid_8^{21} &=& \int_8^{21} 24 \left[ \left[1 - \left( 0.00262629 \ {\bar t} - 0.0236366 \right) \right] \ e^{0.161691 ( T - 24.7 )} \frac{}{} \right. \nonumber \\ 
&& \left( c_1 + c_2h + c_3w + c_4wh + c_5h^2 + c_6w^2 \right)  \frac{ p_{\mbox{\scriptsize protected}} }{ p_{\mbox{\scriptsize morsitans protected}} } \nonumber \\ 
&& \ + \ ( 0.00262629 \ {\bar t} - 0.0236366 ) \ \left( e^{0.110268 T - 9.92201}  + 0.000354783 \right) \nonumber \\ 
&& \left. \frac{100 - h}{100} \ \frac{ p_{\mbox{\scriptsize unprotected}} }{ p_{\mbox{\scriptsize morsitans unprotected}} } \right] \frac{ s_{\mbox{\scriptsize species}} }{ s_{\mbox{\scriptsize morsitans}} } \ \frac{dt}{d{\bar t}} \ d{\bar t}. \nonumber
\end{eqnarray}

\subsubsection*{The Period ${\bar t} = ( 21, 29 ]$}

Transpiration begins its return to unprotected rates during the pharate adult phase. Generalising the exponential growth in Eq. \ref{period21to29} to all species and integrating leads to
\begin{eqnarray} \label{18}
k\mid_{21}^{29} &=& \int_{21}^{29} 24 \left[ \left[ 1 - \left( 6.08165 \times 10^{-7} \ e^{0.454697 {\bar t}} + 0.0443087 \right) \right] e^{0.161691 ( T - 24.7 )} \frac{}{} \right. \nonumber \\ 
&& \left( c_1 + c_2h + c_3w + c_4wh + c_5h^2 + c_6w^2 \right) \ \frac{ p_{\mbox{\scriptsize protected}} }{ p_{\mbox{\scriptsize morsitans protected}} }  \nonumber \\ 
&& + \left( 6.08165 \times 10^{-7} \ e^{0.454697 {\bar t}} + 0.0443087 \right) \left( e^{0.110268 T - 9.92201}  + 0.000354783 \right) \nonumber \\ 
&& \left. \frac{100 - h}{100} \ \frac{ p_{\mbox{\scriptsize unprotected}} }{ p_{\mbox{\scriptsize morsitans unprotected}} } \right] \frac{ s_{\mbox{\scriptsize species}} }{ s_{\mbox{\scriptsize morsitans}} } \ \frac{dt}{d{\bar t}} \ d{\bar t}. \nonumber
\end{eqnarray}

\subsubsection*{The Period ${\bar t} = ( 29, 30 ]$}

There is a return to unprotected rates shortly before eclosion and Eq. \ref{102} once again becomes the applicable integrand, that is 
\begin{eqnarray} \label{114}
k\mid_{29}^{30} &=& \int_{29}^{30} 24 \left( e^{0.110268 T - 9.92201}  + 0.000354783 \right) \frac{100 - h}{100} \ \frac{ p_{\mbox{\scriptsize unprotected}} }{ p_{\mbox{\scriptsize morsitans unprotected}} } \frac{ s_{\mbox{\scriptsize species}} }{ s_{\mbox{\scriptsize morsitans}} } \ \frac{dt}{d{\bar t}} \ d{\bar t}. \nonumber 
\end{eqnarray}

\subsection{A Discrete Function of Time and its Derivative} \label{discreteTimeAndJacobian}

If discrete values are used in Eq. \ref{integrateDerivative} (on page
\pageref{integrateDerivative}), instead of interpolating between the $\tau$
predicted by daily temperature, the integral becomes a sum whose final term is
usually some fraction of a day's contribution,
\begin{eqnarray*} 
\sum_{i=1}^{{\mathop{\rm floor}}\left\{ t \left( {\bar t} \right) \right\}}
\frac{1}{\frac{dt}{d{\bar t}}(i)} \ + \ \frac{ t - \mathop{\rm floor}\left\{ t \right\} }{\frac{dt}{d{\bar t}}({{\mathop{\rm floor}} \left\{ t \right\} + 1})} &=& {\bar t}.
\end{eqnarray*} 
The expression, itself, is suggestive of the method of solution. A progressively
increasing number of days' contributions are summed, until the contributing
interval, itself, overtakes the sum. Newton's method is then implemented to
determine what fraction of the last day is necessary. 

\subsection{Numerical Integration}

Issues of non-differentiability and discontinuity dictate that a low order
integration rule be used. The preferred choice in Childs\nocite{Childs2} (2009)
was Euler's method. Euler's method is, however, no longer appropriate as the
time-dependence is now exponential (`stiff') over parts of the domain. The
midpoint rule is as distasteful from the point of view of its error. The local
error per step, of length ${\Delta t}$, is $\mathop{\rm O}({\Delta t}^3)$. Since
the required number of steps is proportional to $\frac{1}{{\Delta t}}$, the
global error is $\mathop{\rm O}({\Delta t}^2)$. This is indeed primitive. The
real strength of the midpoint rule and other low order methods lies in their
robustness at discontinuities and points of non-differentiability. The maximum,
additional error introduced at such points is of a similar order to the method's
global error. The same cannot be said for the higher order methods. Using the
midpoint rule, one has one problem to solve, whereas using one of the higher
order methods entails solving five, separate problems; each confined to its own
respective domain of Lipshitz continuity etc. The handicap of a poor error is
easily overcome computationally by using a small step length. When considering
the original pupal material used, the known error in the data, that an
engineering-type accuracy is anticipated from the model and a host of other
factors, two significant figures are more than what are sought. Since the
problem is not intractably large, expedience takes precedence over taste and the
more pedestrian midpoint rule is considered the appropriate choice. 

\section{Survival to Eclosion}

How does one translate cumulative water loss into survival? Buxton and
Lewis\nocite{BuxtonAndLewis1} (1934) and Bursell\nocite{Bursell1} (1958)
collected pupal emergence data for a variety of species, over a range of
humidities at 24~$^\circ\mathrm{C}$ (all reported in Bursell\nocite{Bursell1},
1958). Two challenges arise in using this pupal emergence data. The first is to
establish a credible relationship between the proportion which eclose and the
humidity of the substrate, while the second is how to relate this survival to
total water loss when survival is only known as a function of humidity at
24~$^\circ\mathrm{C}$. Of course, the former problem reduces to an exercise in
curve-fitting, once a suitable function has been determined.

\begin{figure}[H]
    \begin{center}
\includegraphics[height=13cm, angle=-90, clip = true]{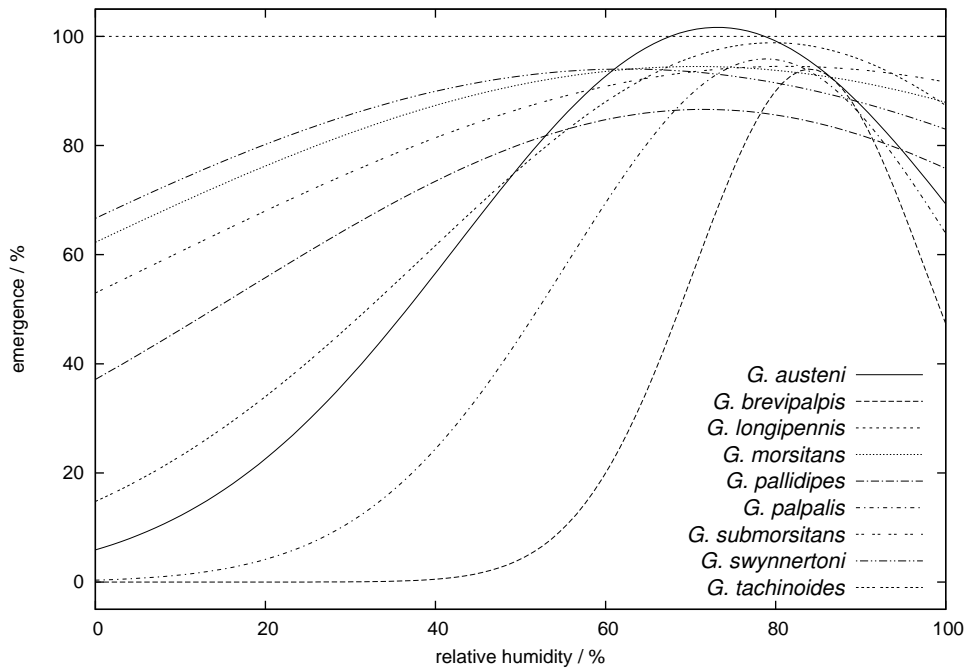}
\caption{Percentage emergence data modelled as a Gaussian curve
(Childs\protect\nocite{Childs2}, 2009) for a variety of species. All are at 24~$^\circ\mathrm{C}$, except {\em G. tachinoides} (30~$^\circ\mathrm{C}$). A
straight line had to be fitted to the only two data points for the single
exception, {\em G. longipennis}} \label{allSpeciesTogether}
   \end{center}
\end{figure}

What is the relationship between pupal emergence and humidity in Fig. 12 of
Bursell\nocite{Bursell1} (1958)? Assuming the usual intra-specific variation,
the simplest point of departure is that some pupae will be slightly bigger, have
slightly bigger reserves and more competent integuments. Yet others will be
slightly smaller, have slightly smaller reserves and less competent integuments.
This justifies the following interpretation of the Bursell\nocite{Bursell1}
(1958) Fig. 12 data.
\begin{assumption} \label{assumption4} 
{\bf \em The relationship between pupal emergence and humidity is a Gaussian curve, or a part thereof.} 
\end{assumption} 
The parameters in 
\begin{eqnarray}
E(h_{\mbox{\scriptsize 24}}) &=& a \ \mathop{\rm exp}\left[ - \frac{(h_{\mbox{\scriptsize 24}} - b)^2}{2 c^2} \right], \nonumber
\end{eqnarray}
for each species, are provided in Table \ref{emergence}. Convincing fits to the
data are obtained in this way, regardless of whether or not the underlying logic
is correct (Fig. \ref{allSpeciesTogether} and Table \ref{emergence}). 

A similar challenge to that encountered for historically-conditioned
transpiration, compounds matters when it comes to the survival to eclosion for
each species. Humidities are steady, furthermore, the data were obtained at a
constant 24~$^\circ\mathrm{C}$. 
\begin{assumption} \label{assumption5} 
{\bf \em The pupal survival to eclosion, for a given water loss, is the same as that for the steady humidity, at 24~$^\circ\mathrm{C}$, that produced an equivalent total water loss.} 
\end{assumption} 
In other words, it is assumed that survival to eclosion can be re-expressed in
terms of total water loss. Do different histories in temperature and humidity,
which produce the same water loss, imply the same pupal emergence, or is the
level of the organism's water reserve at some particular stage more relevant to
the pupa's survival to full term? Such dependence is beyond the scope of this
research, as well as that of Bursell\nocite{Bursell1} (1958). In practice, it is
far easier to convert a total computed water loss to a corresponding
steady-humidity-at-24~$^\circ\mathrm{C}$, instead of the other way around. The
problem is that the water loss algorithm is relatively involved and voluminous.
The results of the water loss algorithm at 24~$^\circ\mathrm{C}$ constitute a
monotonic decline with steady humidity, barring excretion. These are ideal
circumstances for the implementation of a half interval search (the rate of
convergence is not bad in this instance).
\begin{table}[H]
\begin{center}
\begin{tabular}{l l | c c c}  
&  &  &  & \\
~ ~ group & ~ species & \ \ $a$ & $b$ & $c$ \\ 
&  &  &  & \\ \hline 
&  &  &  & \\
{\em morsitans} & {\em austeni} & $101.663$ & $73.1591$ & $30.6468$ \\ 
 \ & {\em morsitans} & \ \  $94.4792$ & $70.6391$ & $77.3495$ \\ 
 \ & {\em pallidipes} & \ \ $86.6257$ & $71.5636$ & $54.9713$ \\ 
 \ & {\em submorsitans} & \ \ $94.5092$ & $81.1895$ & $75.4474$ \\ 
 \ & {\em swynnertoni} & \ \ $ 94.0194$ & $62.4064$ & $75.2339$ \\ 
&  &  &  & \\
{\em palpalis} & {\em palpalis} & \ \ $95.8732$ & $78.8419$ & $23.4835$ \\
 \ & {\em tachinoides} & \ \ $98.8383$ & $79.6877$ & $40.8616$ \\ 
&  &  &  & \\
{\em fusca} & {\em brevipalpis} \ & \ \ $94.0057$ & $84.0199$ & $13.6433$ \\ 
&  &  &  & \\
\end{tabular}
\caption{Parameters for the fit of a Gaussian curve to the
Bursell\nocite{Bursell1} (1958) and Buxton and Lewis\nocite{BuxtonAndLewis1}
(1934) pupal emergence data for a variety of species (Childs, 2009).
All are at \mbox{24 $^\circ\mathrm{C}$}, except {\em G. tachinoides}
(30~$^\circ\mathrm{C}$).} \label{emergence}
\end{center}
\end{table}

\section{Results}

\begin{figure}[H]
   \begin{center}
\includegraphics[height=11cm, angle=0, clip = true]{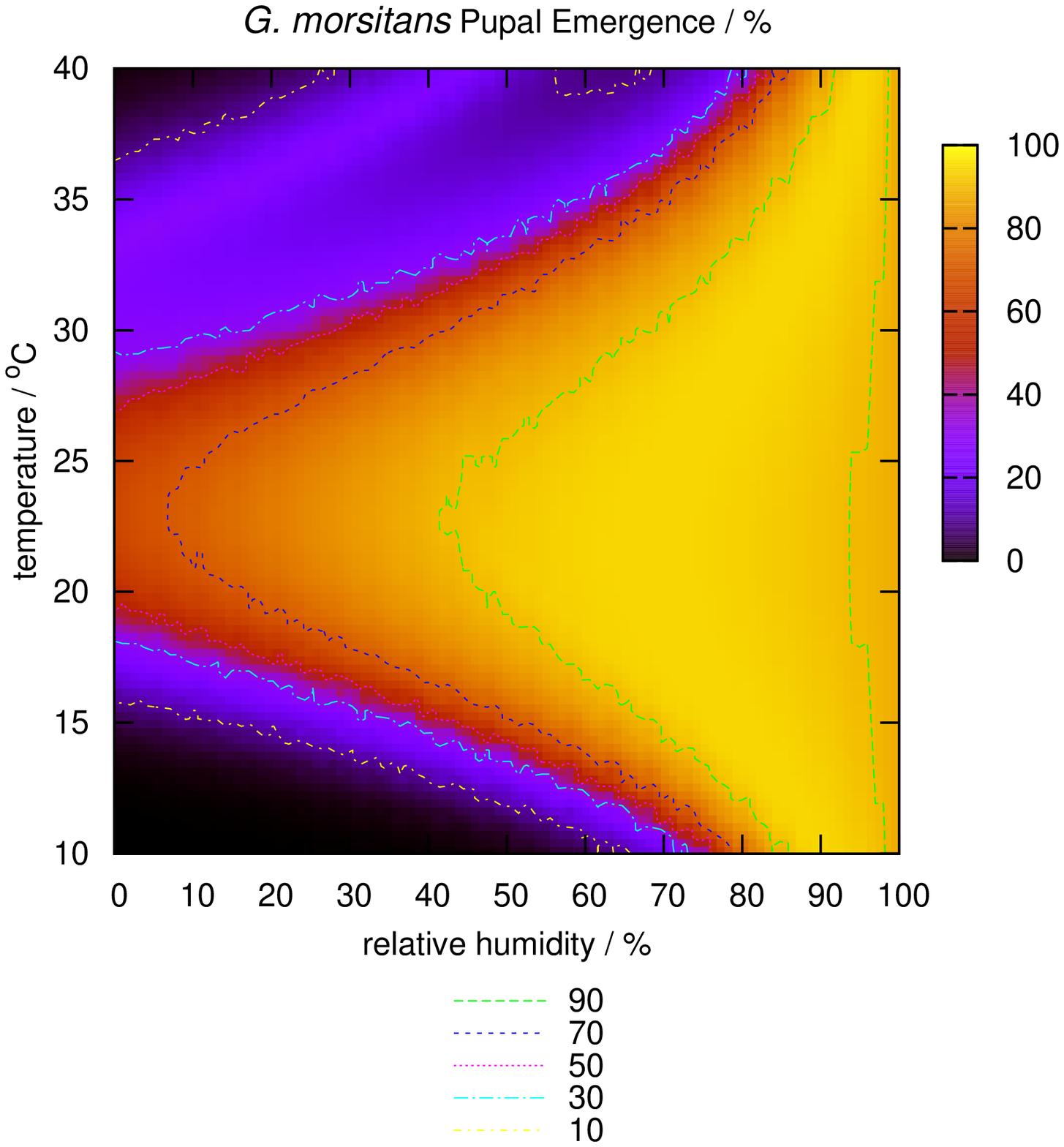}
\includegraphics[height=11cm, angle=0, clip = true]{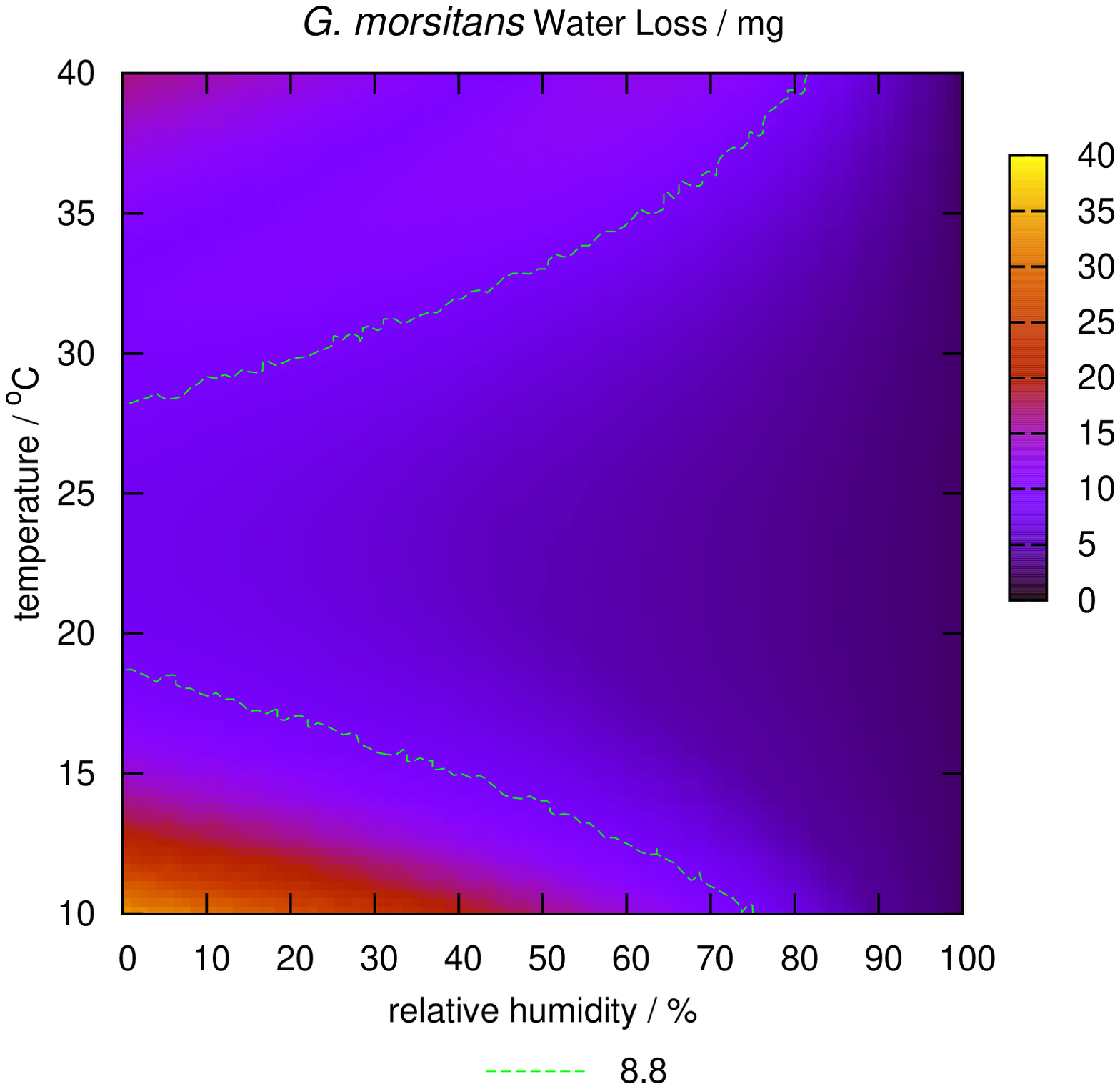}
\includegraphics[height=11cm, angle=0, clip = true]{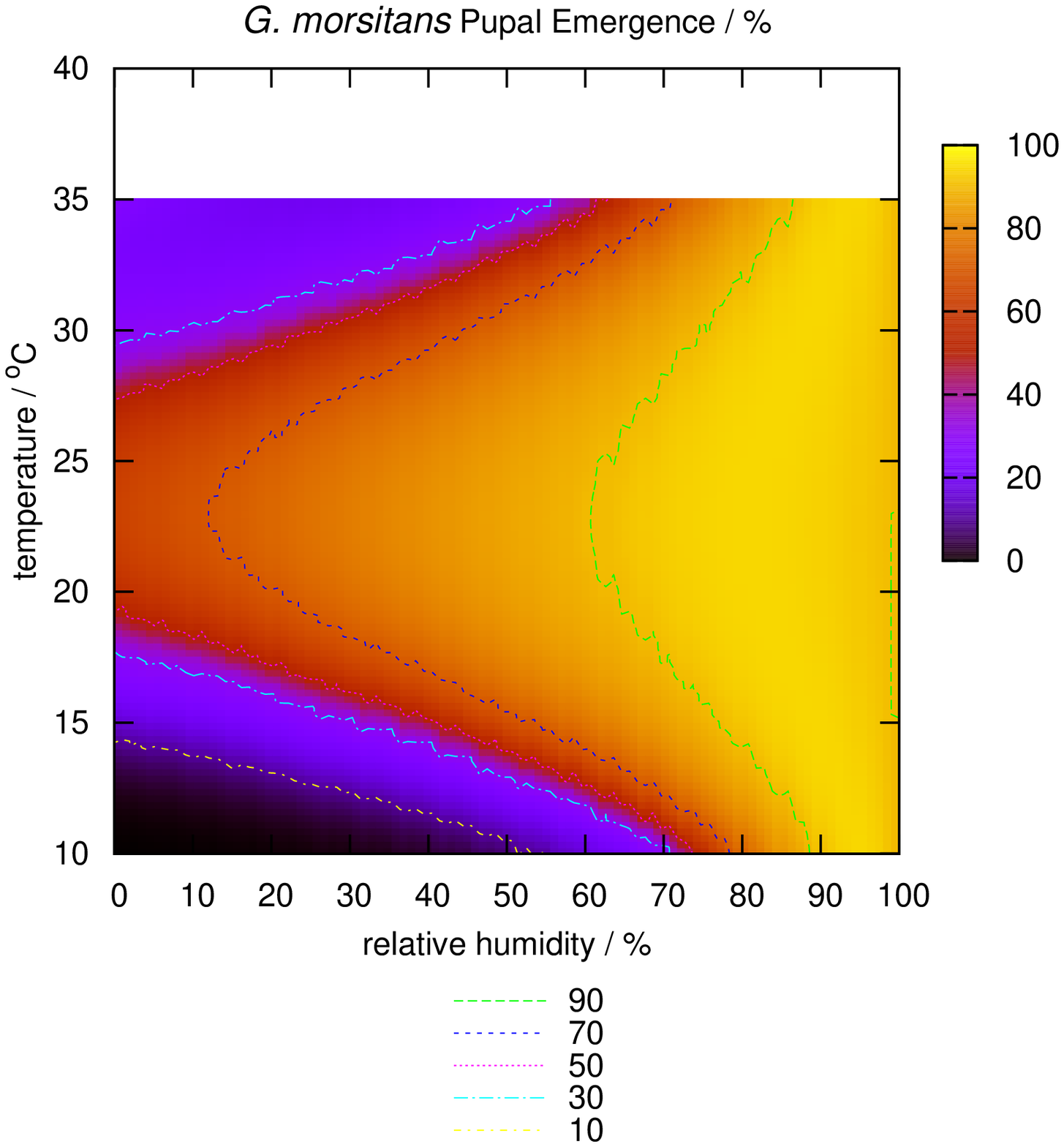}
\includegraphics[height=11cm, angle=0, clip = true]{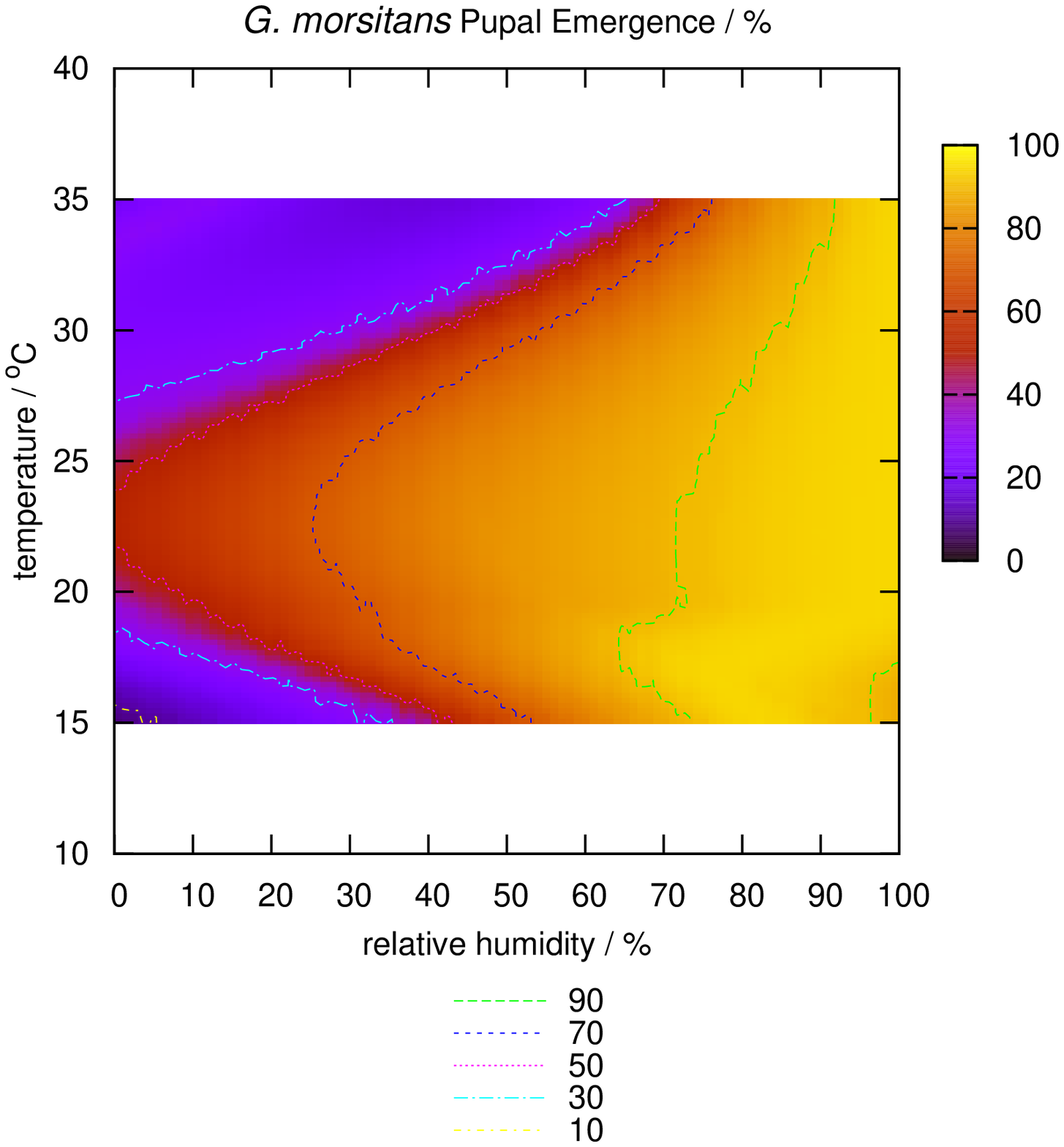}
\caption{{\em G. morsitans} pupal emergence (top left) and water loss (top right); {\em G. morsitans} pupal emergence for a 35~$^\circ\mathrm{C}$ and 25\% $\mathrm{r.h.}$ heat wave on the first two days after larviposition (bottom left) and on day fifteen and sixteen (bottom right).} \label{morsitansHeatWave}
   \end{center}
\end{figure}

\begin{figure}[H]
    \begin{center}
\includegraphics[height=11cm, angle=0, clip = true]{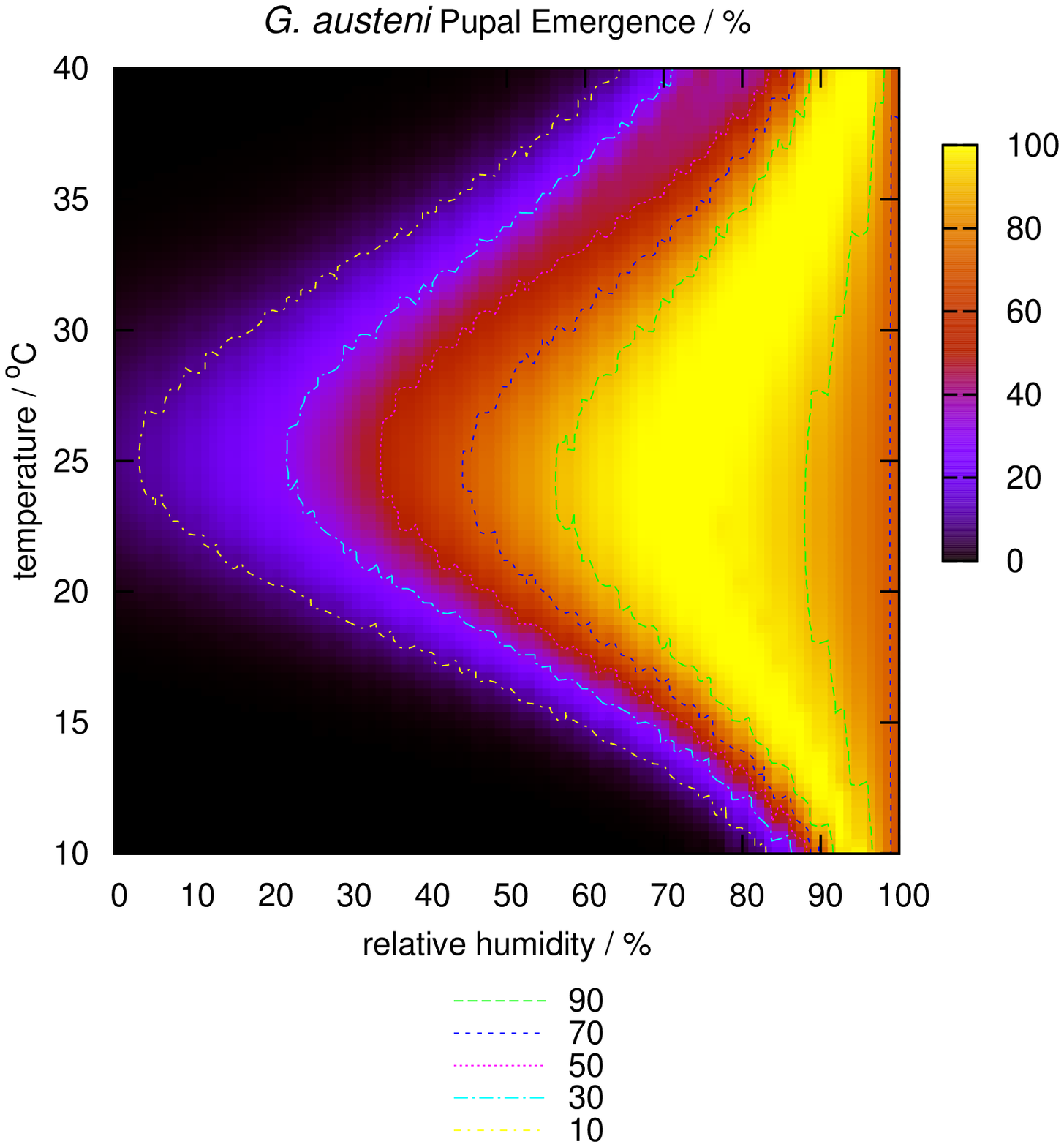}
\includegraphics[height=11cm, angle=0, clip = true]{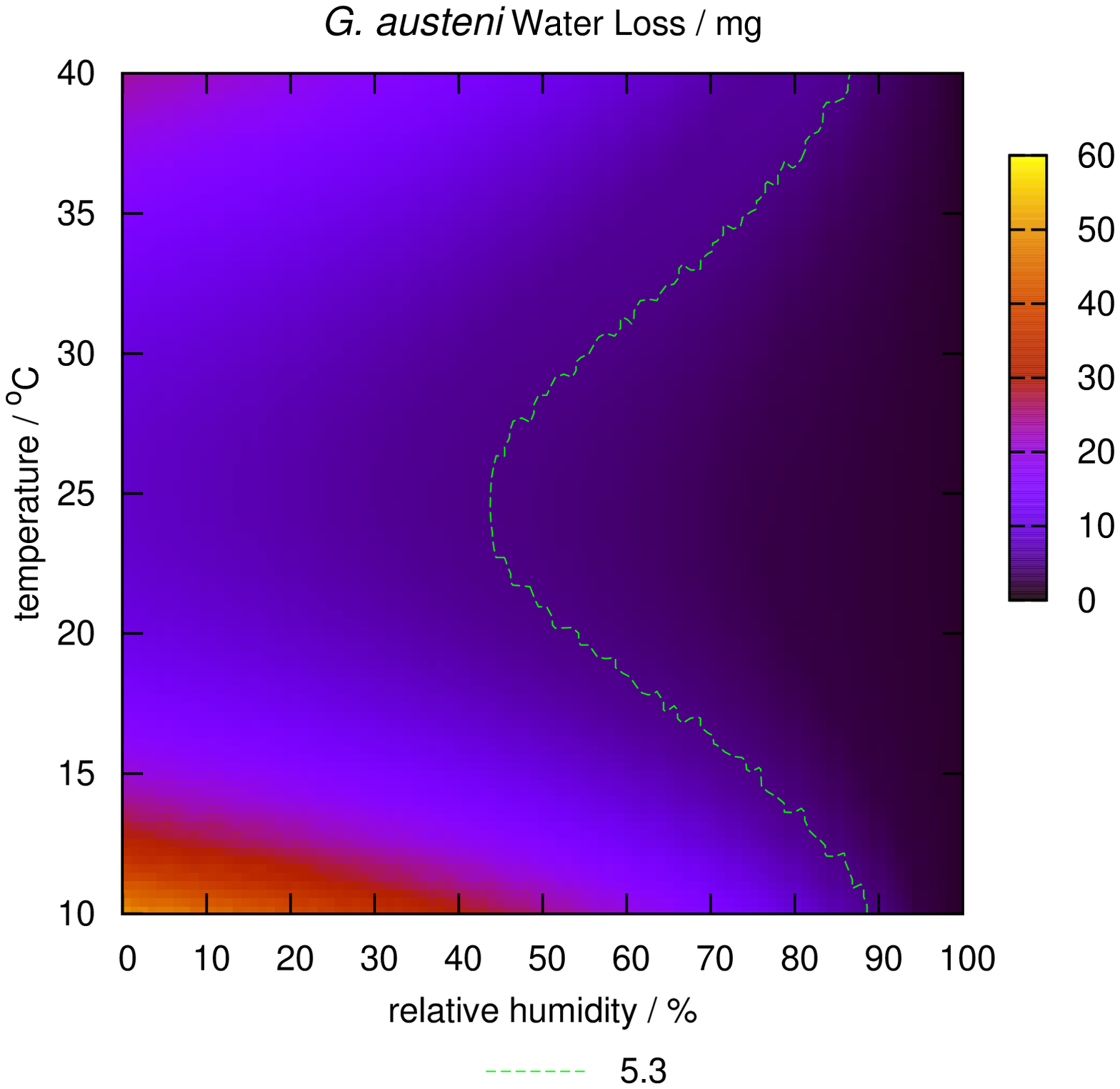}
\includegraphics[height=11cm, angle=0, clip = true]{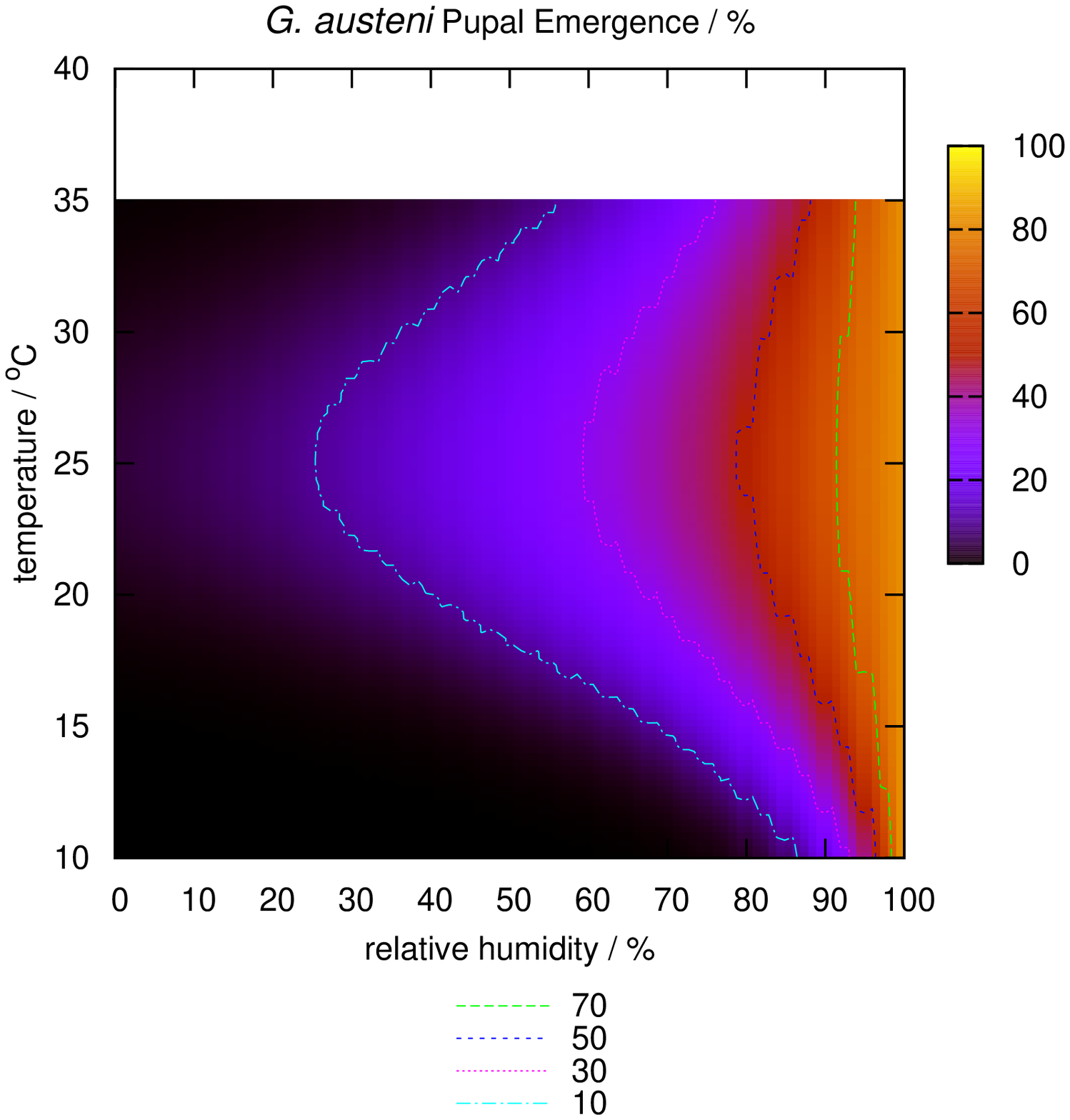}
\includegraphics[height=11cm, angle=0, clip = true]{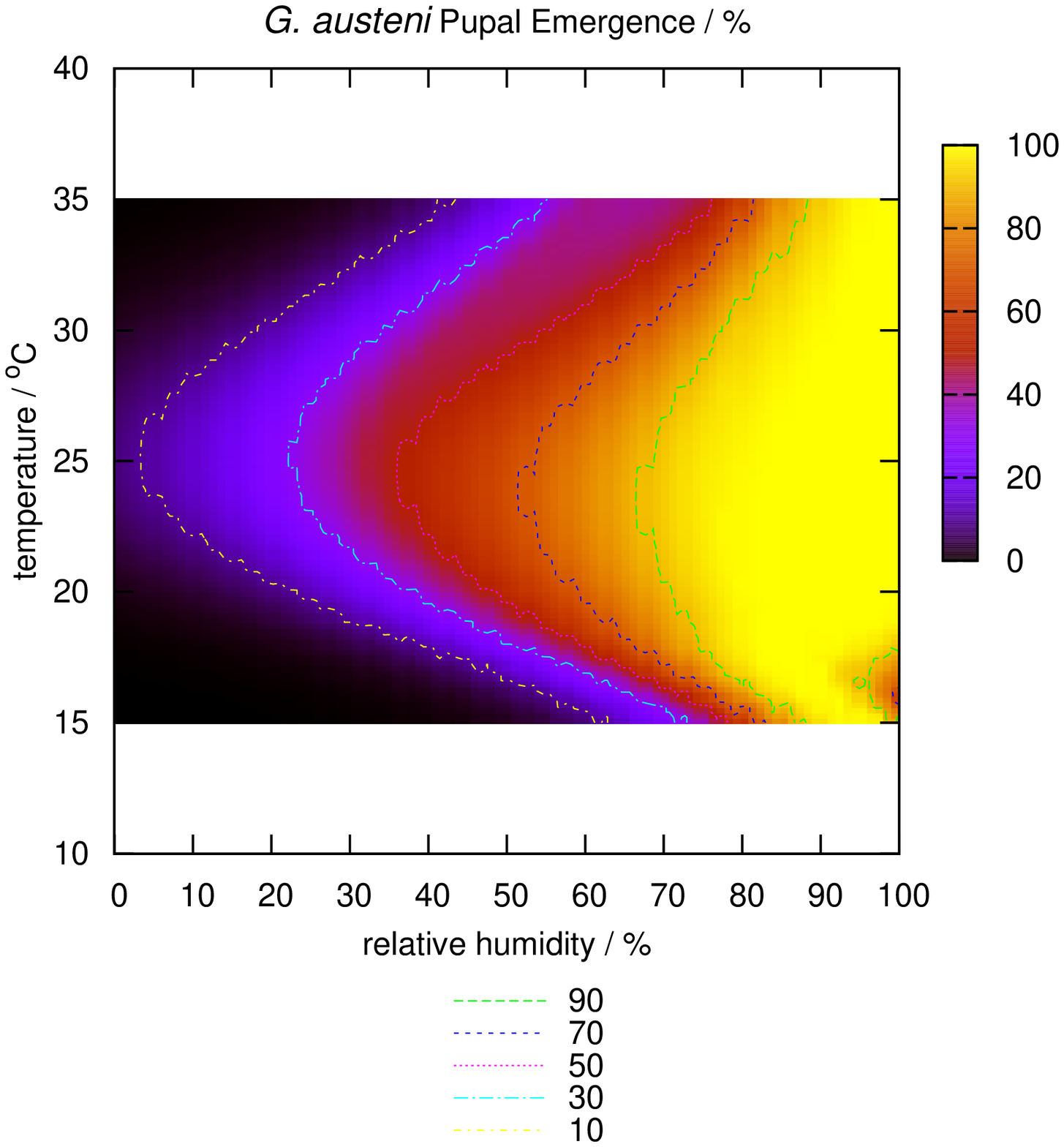}
\caption{{\em G. austeni} pupal emergence (top left) and water loss (top right); {\em G. austeni} pupal emergence for a 35~$^\circ\mathrm{C}$ and 25\% $\mathrm{r.h.}$ heat wave on the first two days after larviposition (bottom left) and on day fifteen and sixteen (bottom right).} \label{austeniHeatWave}
   \end{center}
\end{figure}

\begin{figure}[H]
    \begin{center}
\includegraphics[height=11cm, angle=0, clip = true]{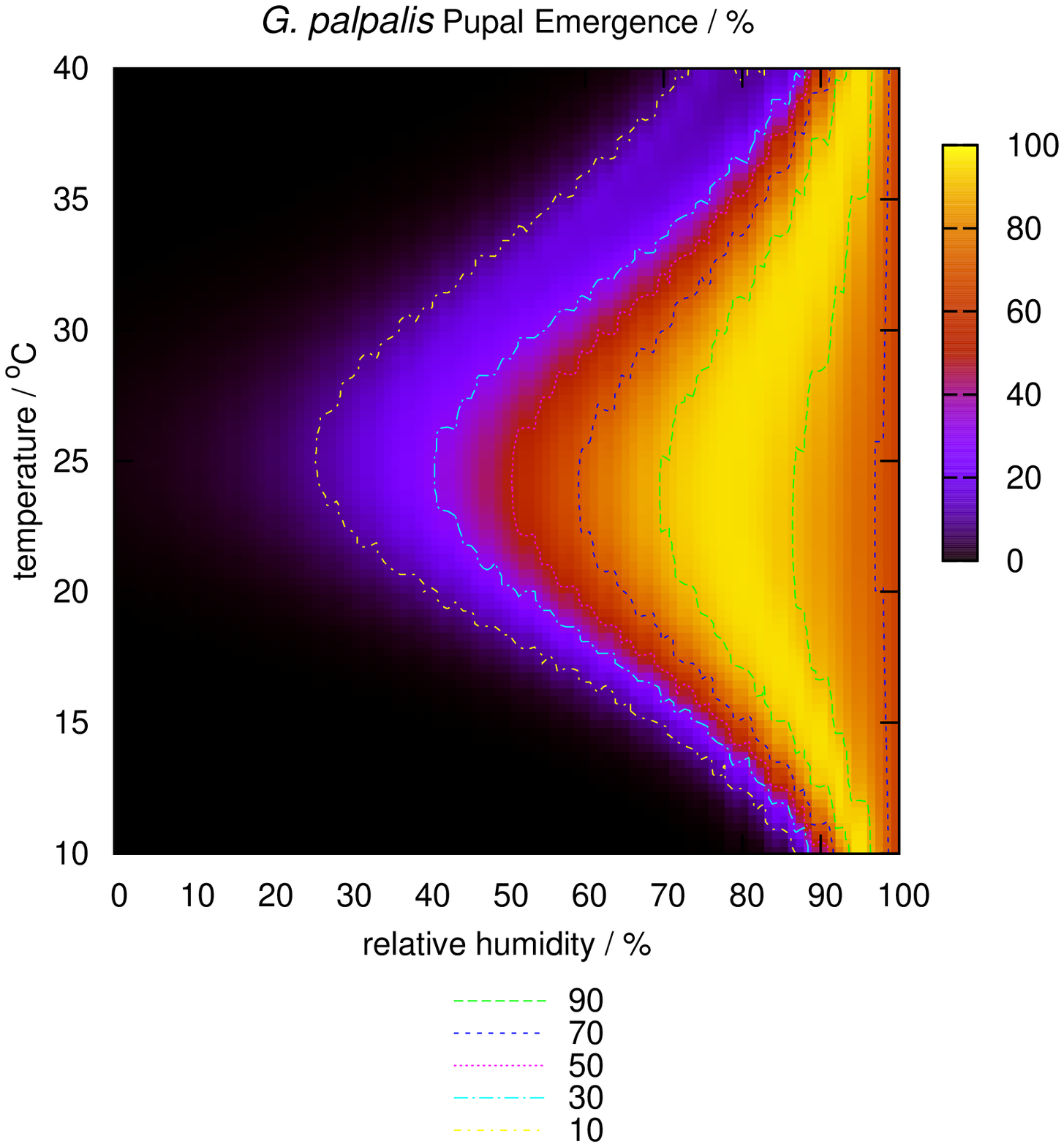}
\includegraphics[height=11cm, angle=0, clip = true]{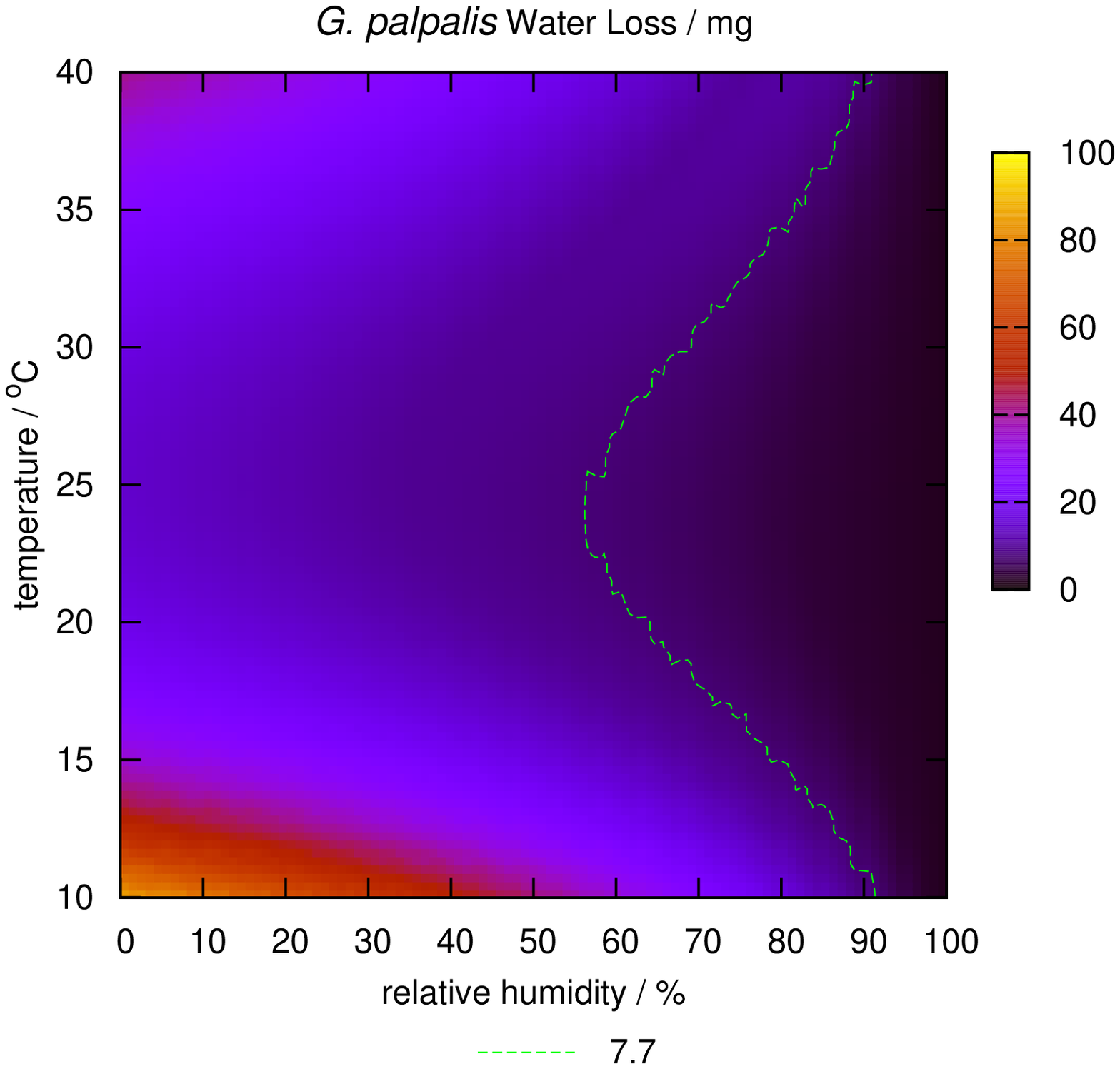}
\includegraphics[height=11cm, angle=0, clip = true]{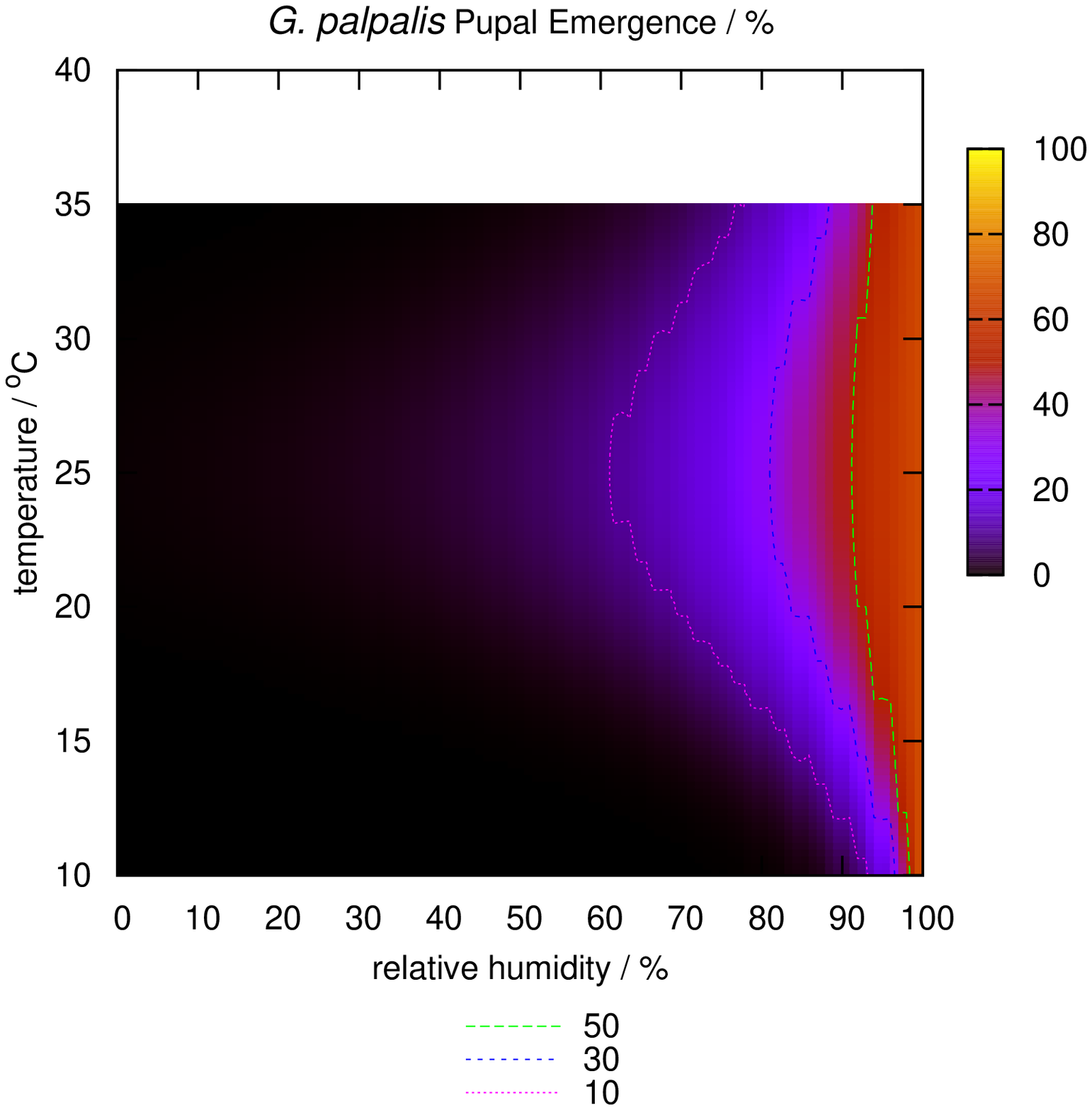}
\includegraphics[height=11cm, angle=0, clip = true]{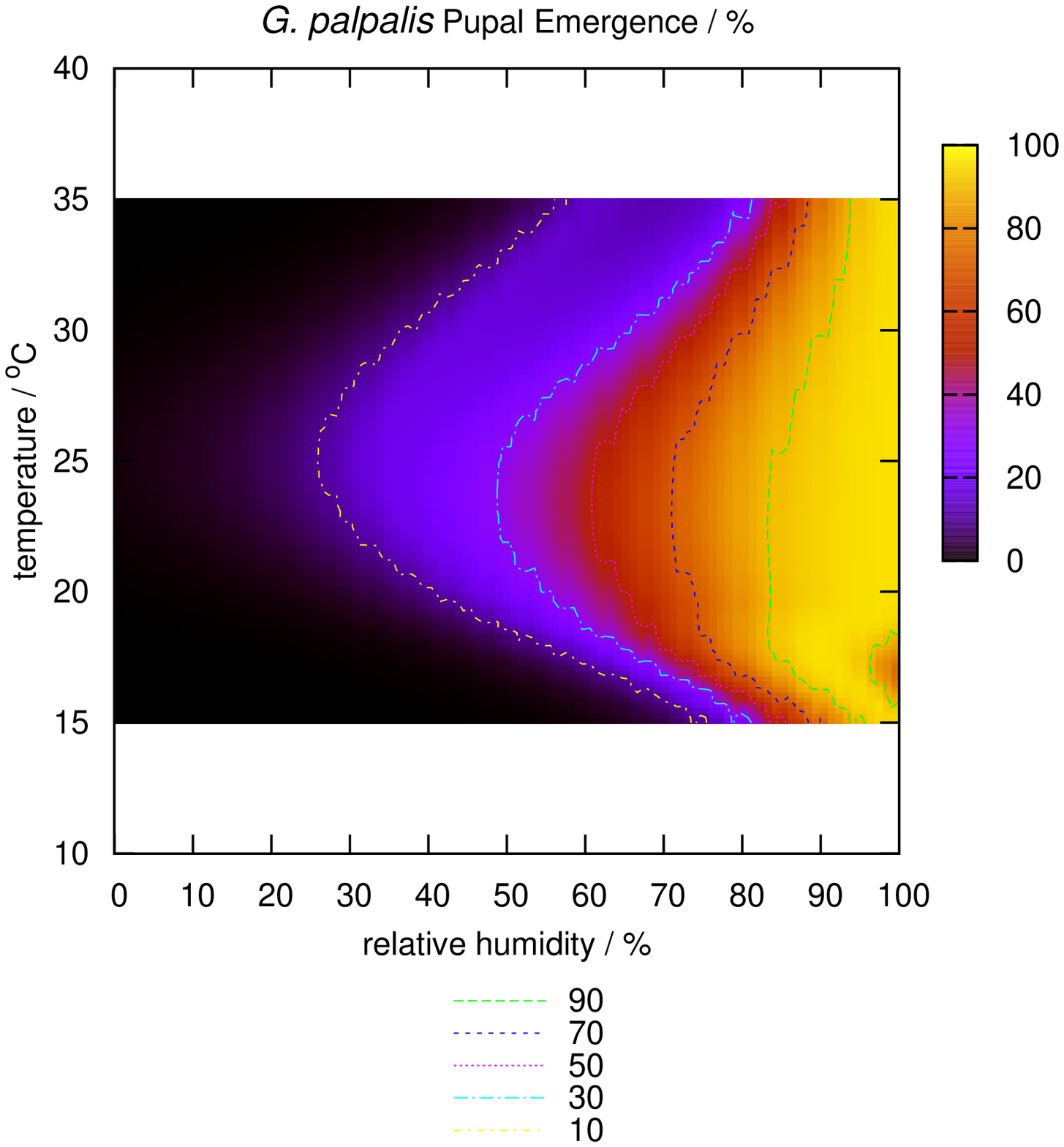}
\caption{{\em G. palpalis} pupal emergence (top left) and water loss (top right); {\em G. palpalis} pupal emergence for a 35~$^\circ\mathrm{C}$ and 25\% $\mathrm{r.h.}$ heat wave on the first two days after larviposition (bottom left) and on day fifteen and sixteen (bottom right).} \label{palpalisHeatWave}
   \end{center}
\end{figure}

\begin{figure}[H]
    \begin{center}
\includegraphics[height=11cm, angle=0, clip = true]{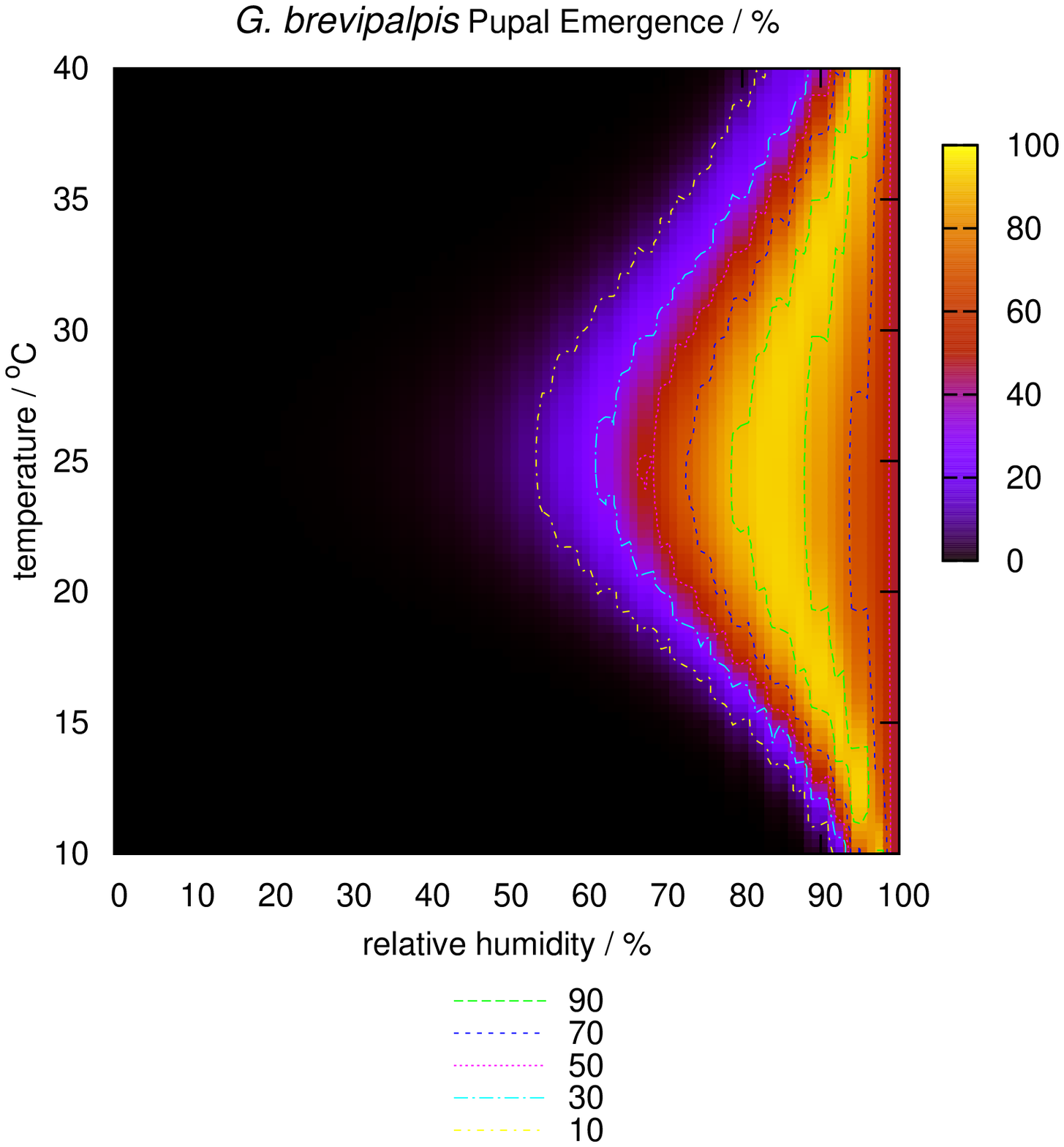}
\includegraphics[height=11cm, angle=0, clip = true]{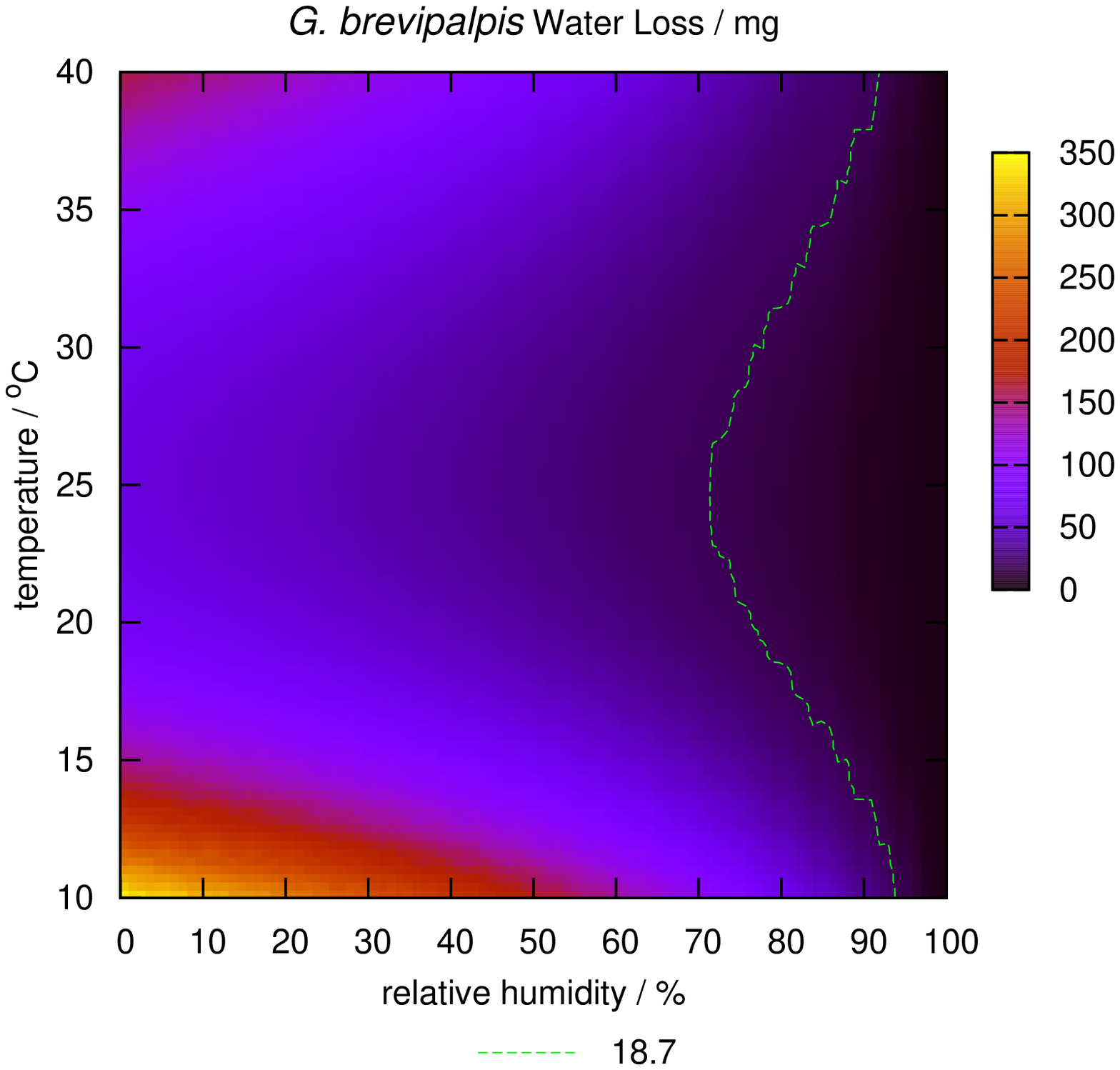}
\includegraphics[height=11cm, angle=0, clip = true]{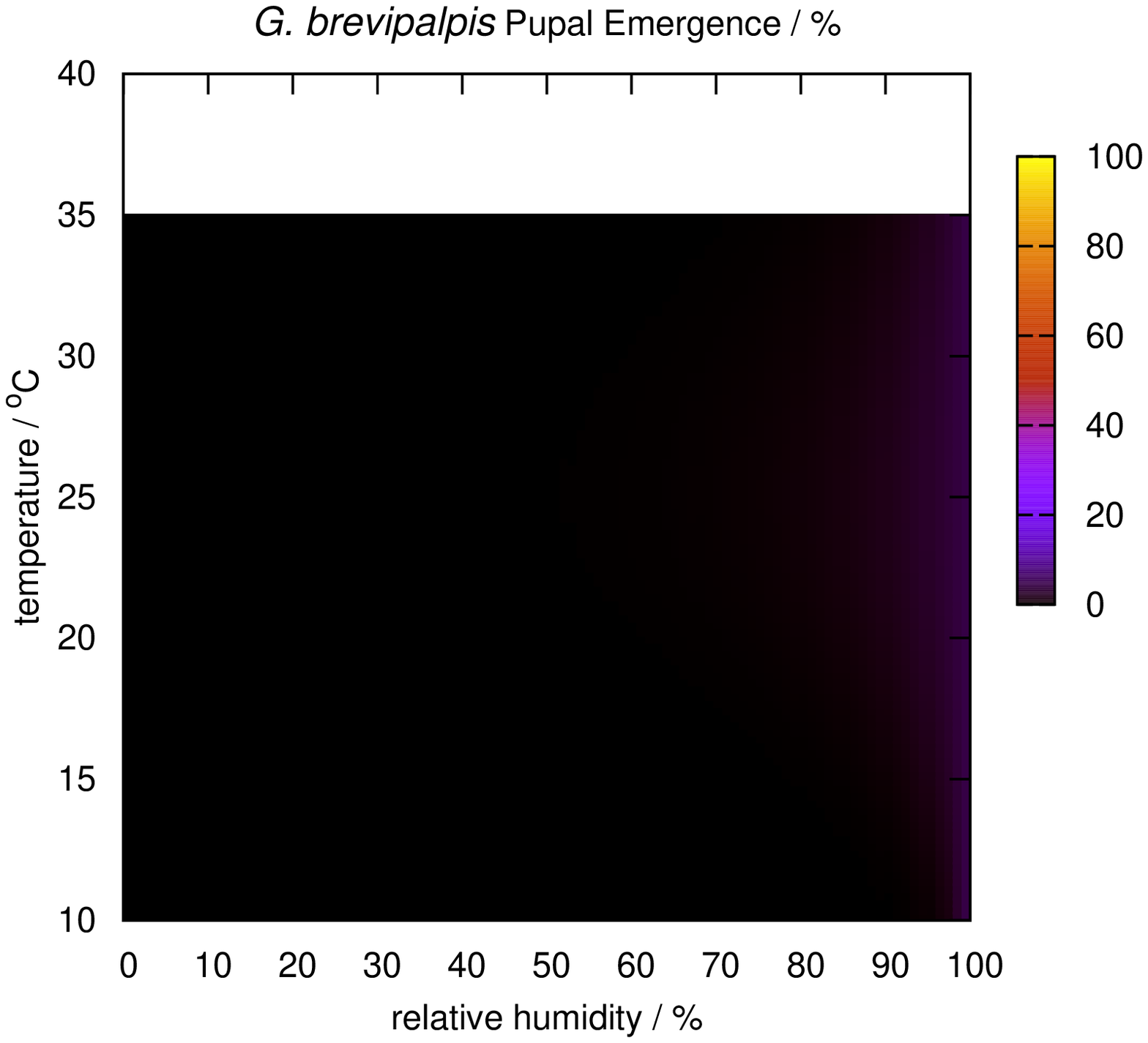}
\includegraphics[height=11cm, angle=0, clip = true]{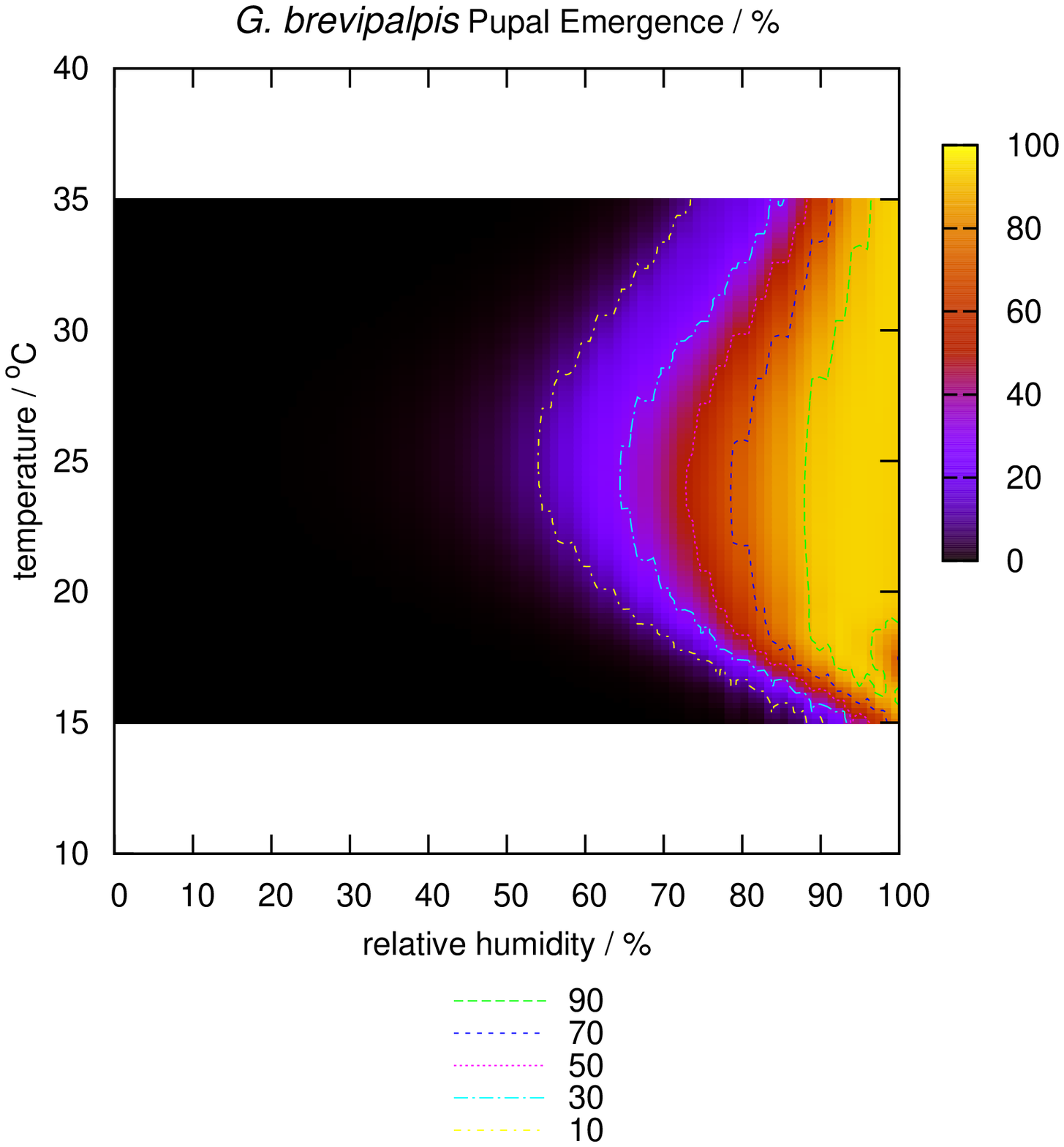}
\caption{{\em G. brevipalpis} pupal emergence (top left) and water loss (top right); {\em G. brevipalpis} pupal emergence for a 35~$^\circ\mathrm{C}$ and 25\% $\mathrm{r.h.}$ heat wave on the first two days after larviposition (bottom left) and on day fifteen and sixteen (bottom right).} \label{brevipalpisHeatWave}
   \end{center}
\end{figure}

\begin{figure}[H]
    \begin{center}
\includegraphics[height=11cm, angle=0, clip = true]{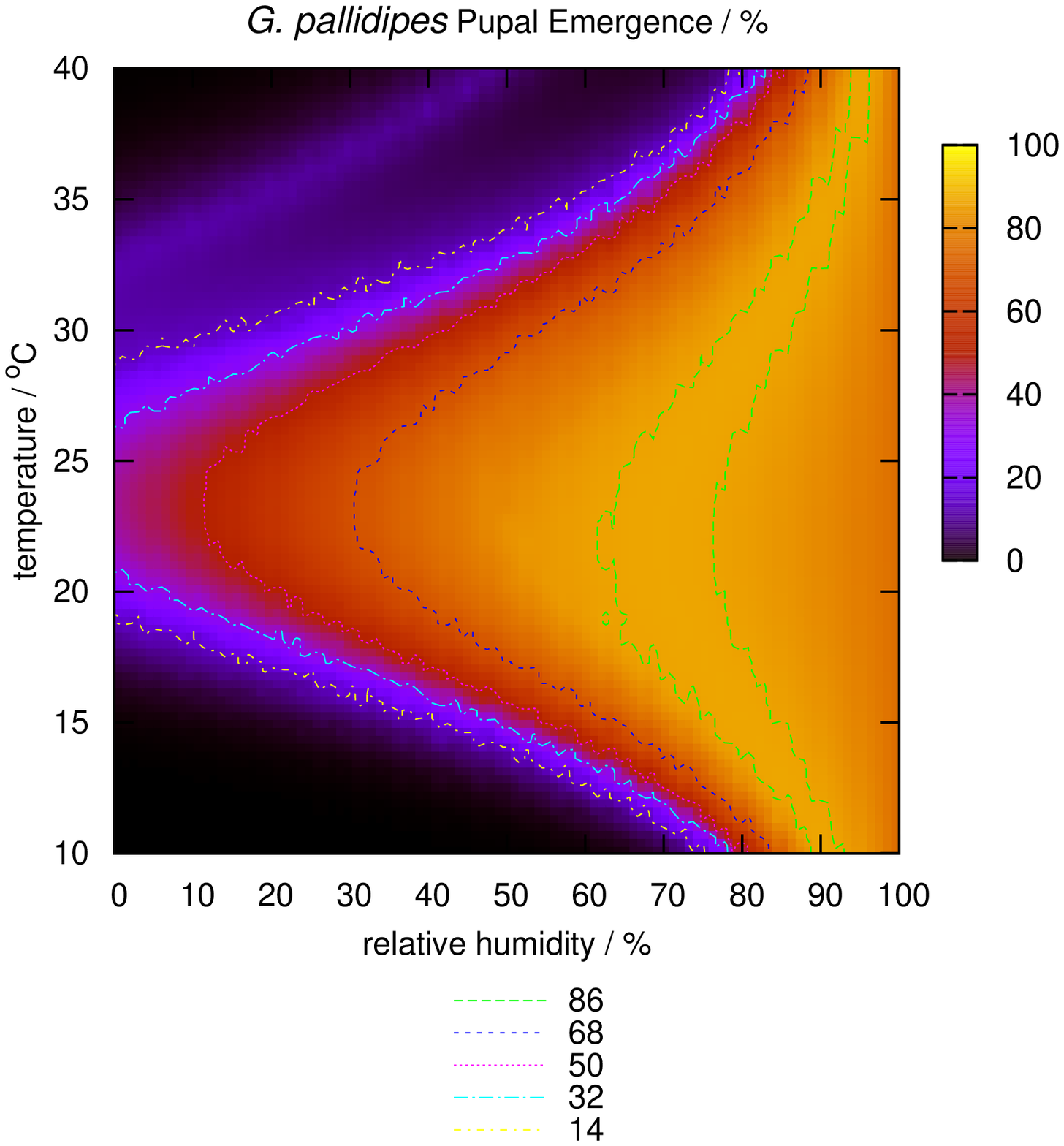}
\includegraphics[height=11cm, angle=0, clip = true]{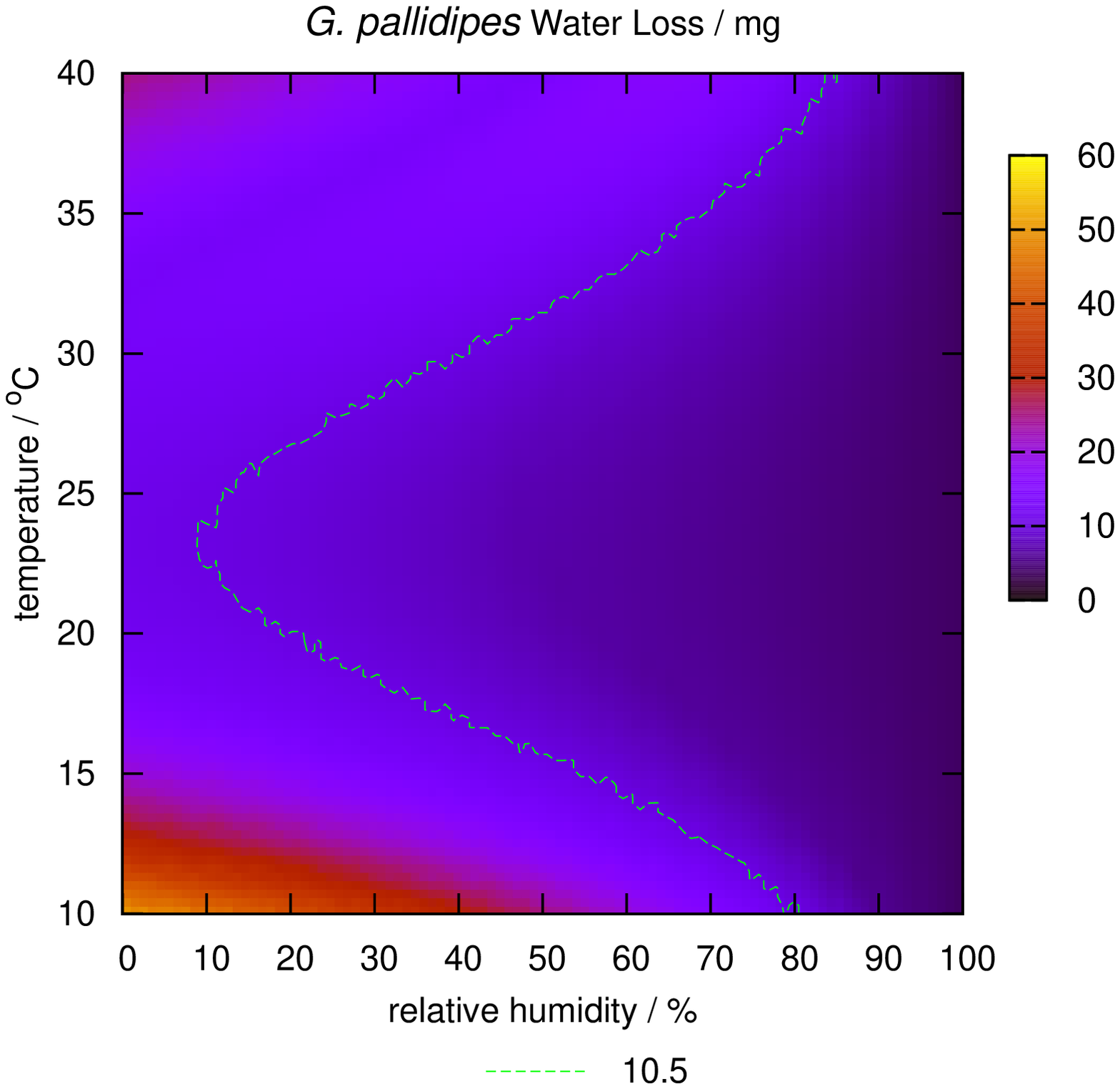}
\includegraphics[height=11cm, angle=0, clip = true]{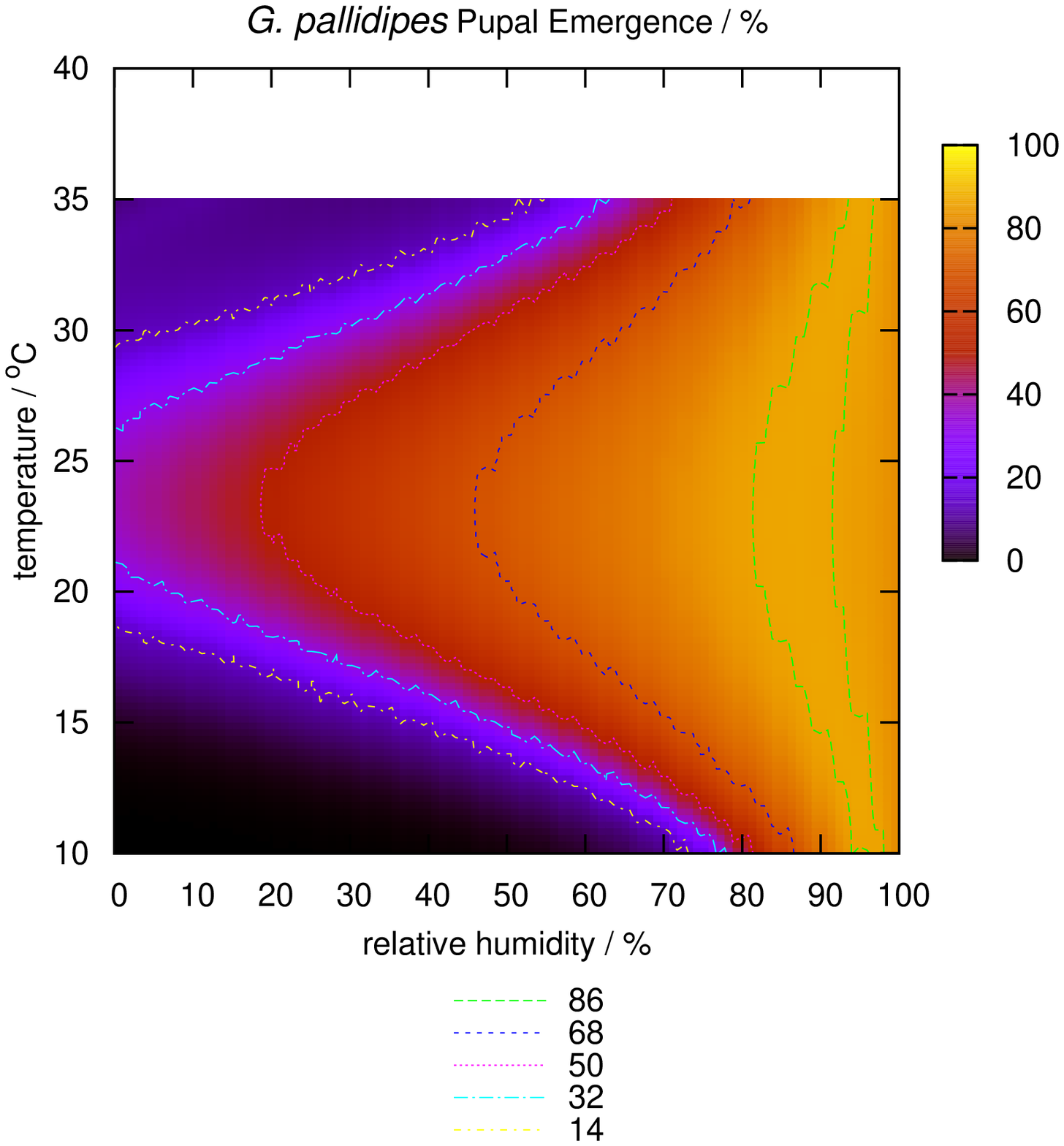}
\includegraphics[height=11cm, angle=0, clip = true]{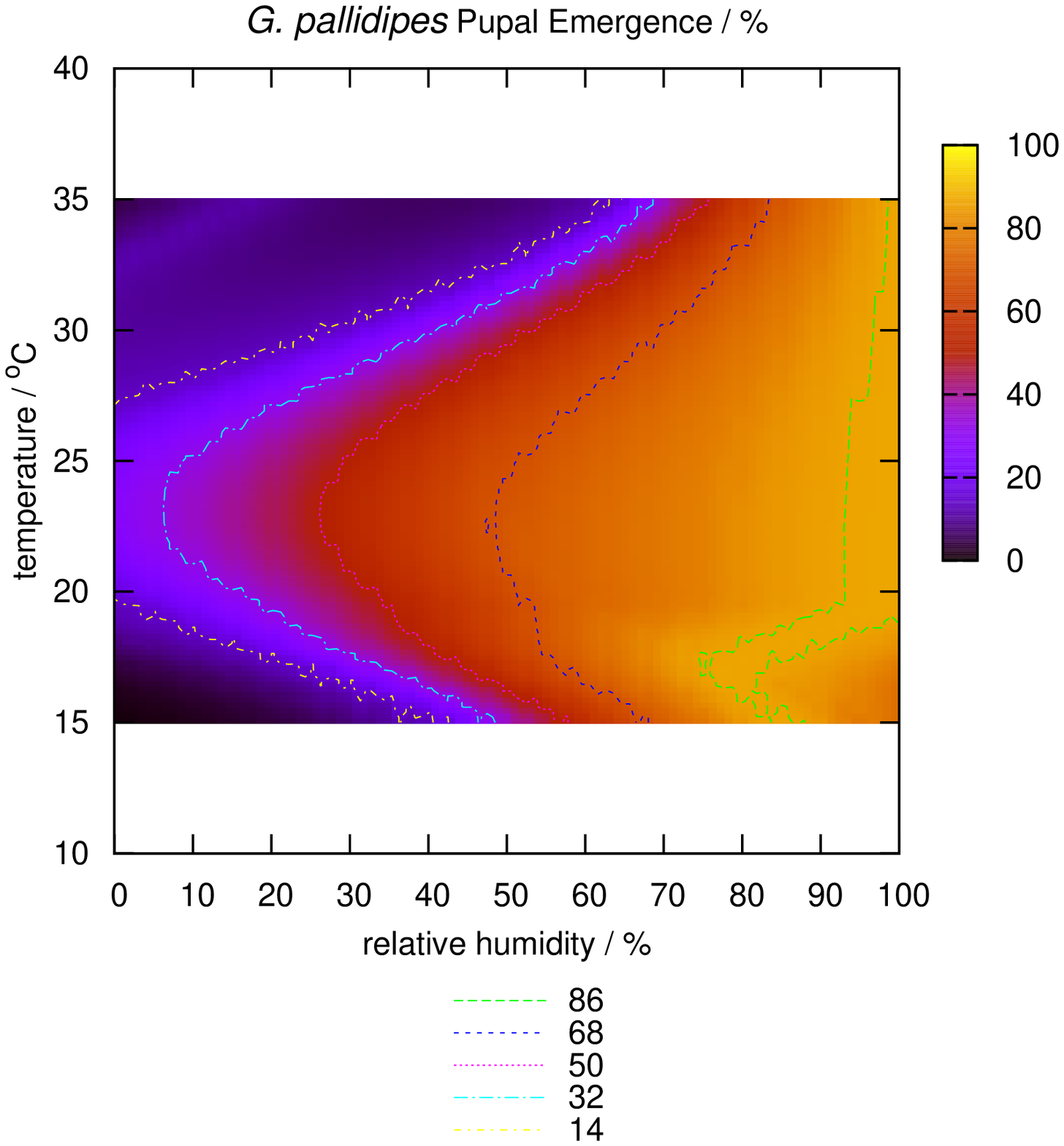}
\caption{{\em G. pallidipes} pupal emergence (top left) and water loss (top right); {\em G. pallidipes} pupal emergence for a 35~$^\circ\mathrm{C}$ and 25\% $\mathrm{r.h.}$ heat wave on the first two days after larviposition (bottom left) and on day fifteen and sixteen (bottom right).} \label{pallidipesHeatWave}
   \end{center}
\end{figure}

\begin{figure}[H]
    \begin{center}
\includegraphics[height=11cm, angle=0, clip = true]{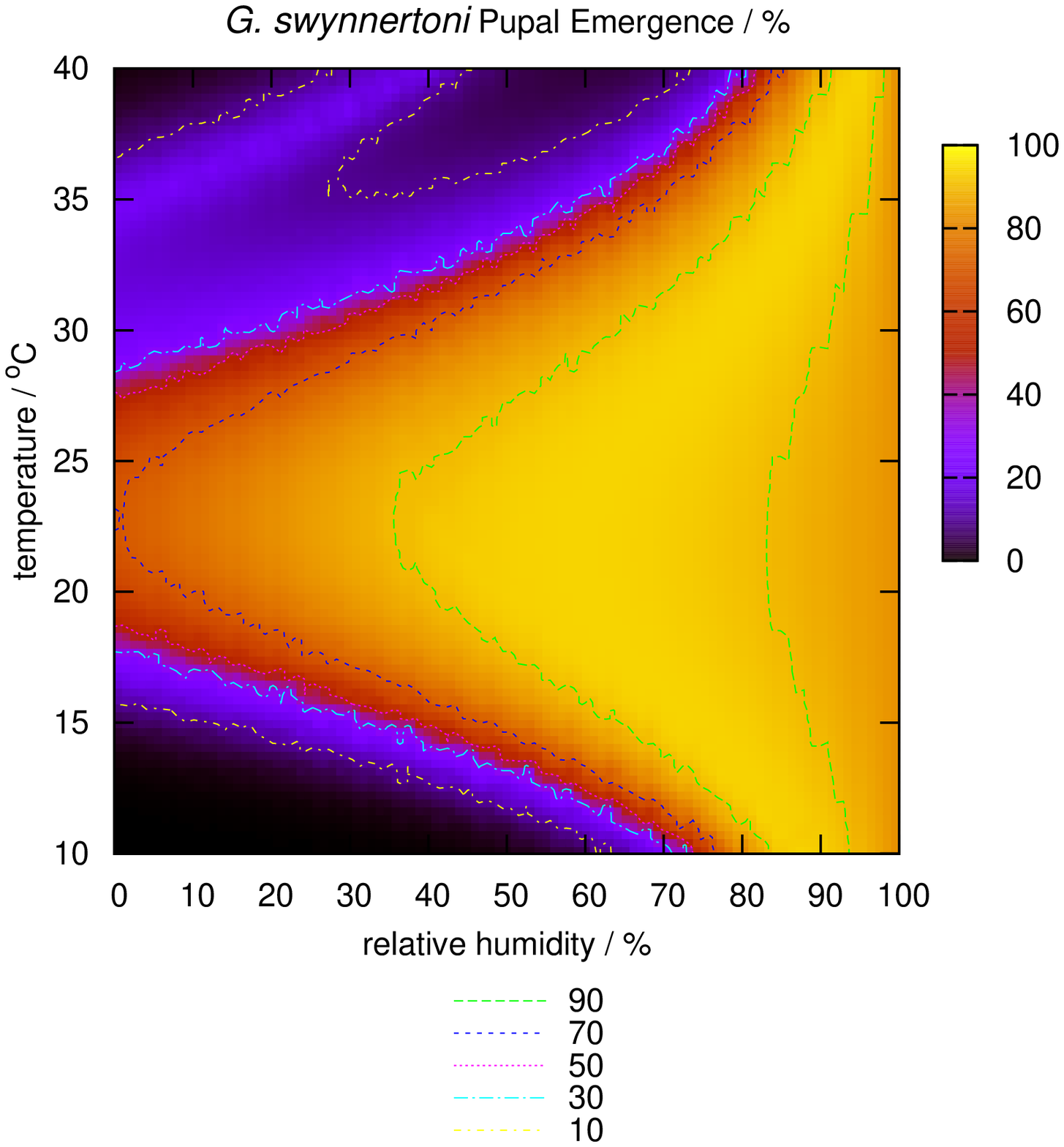}
\includegraphics[height=11cm, angle=0, clip = true]{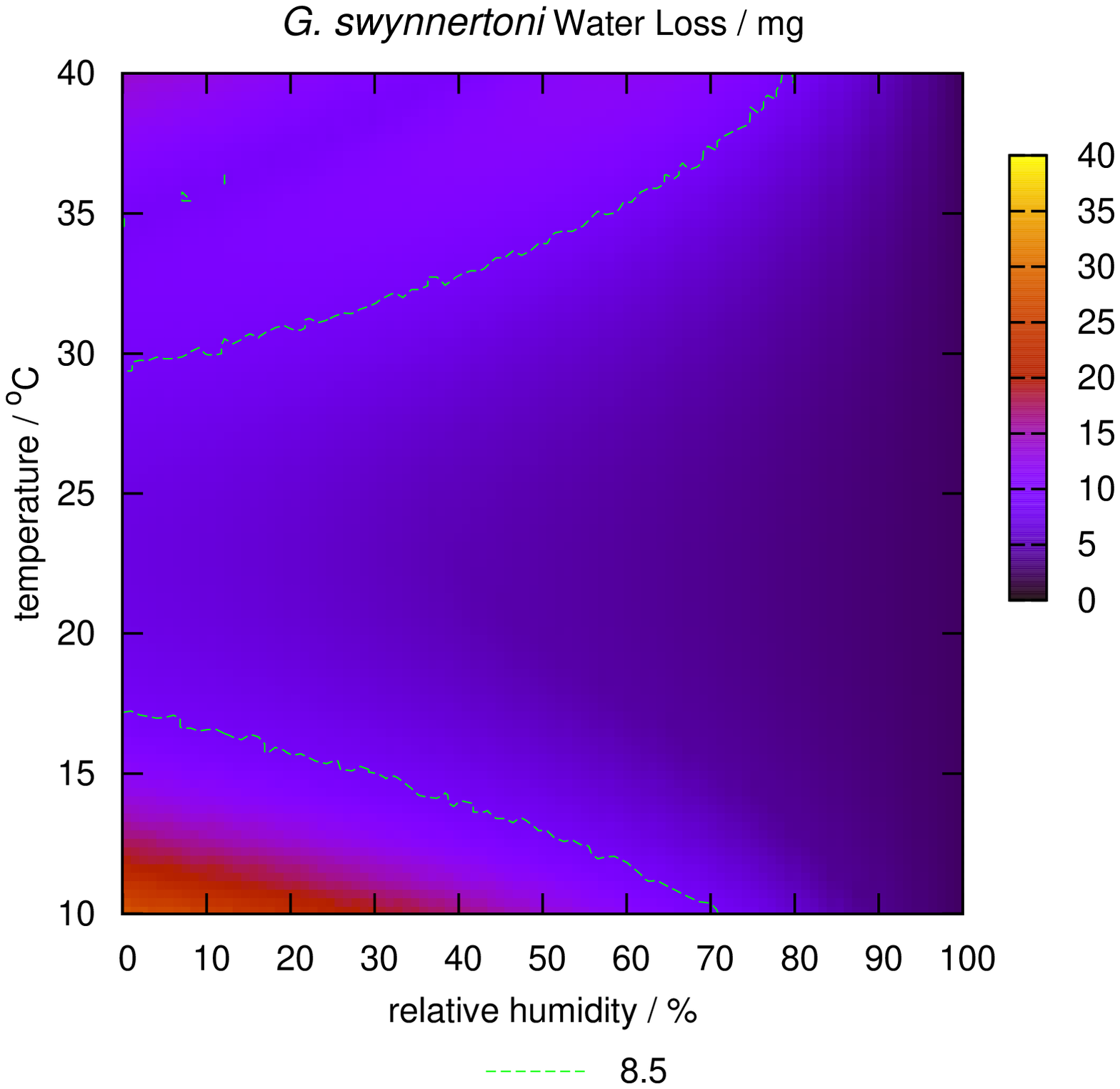}
\includegraphics[height=11cm, angle=0, clip = true]{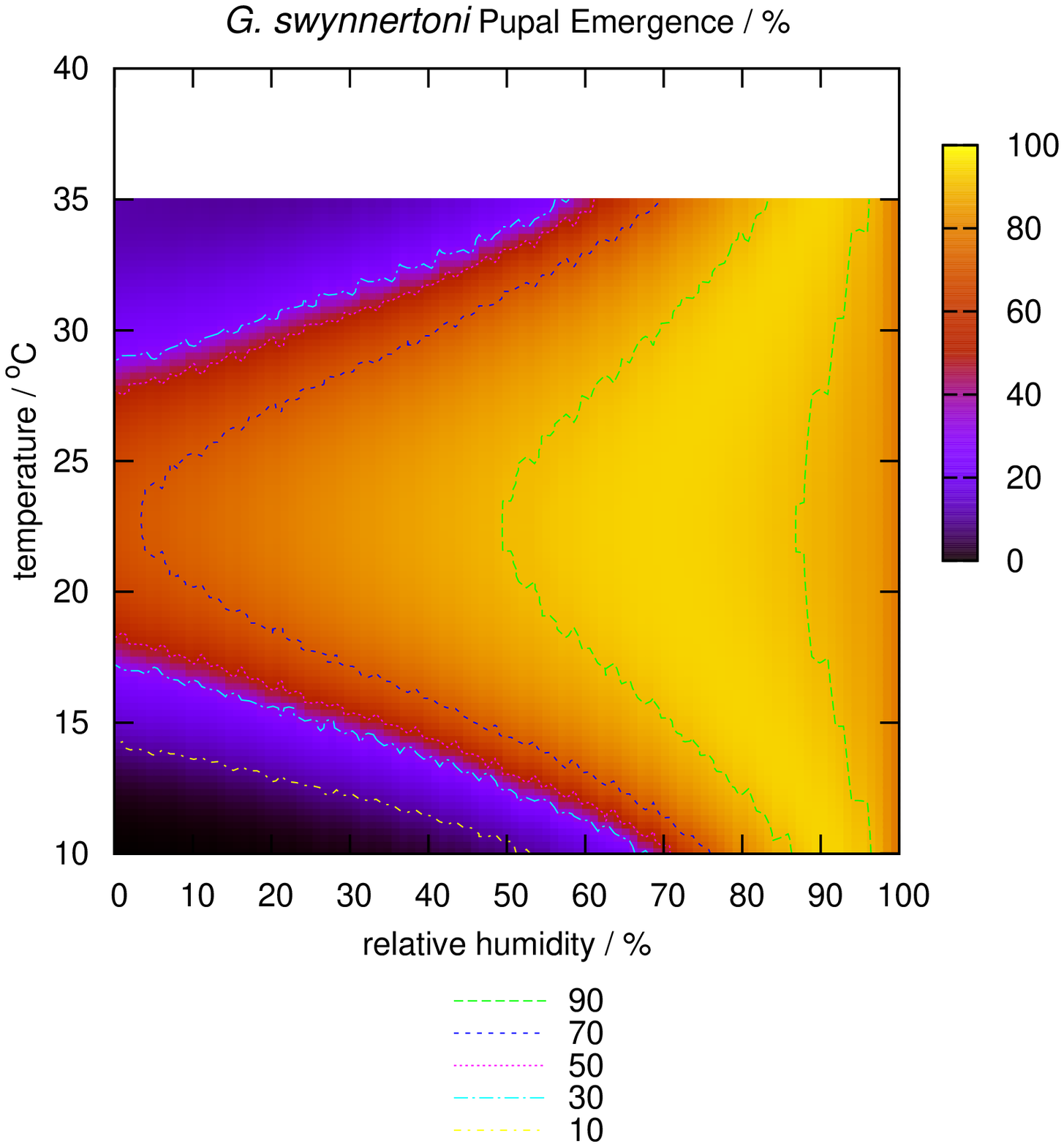}
\includegraphics[height=11cm, angle=0, clip = true]{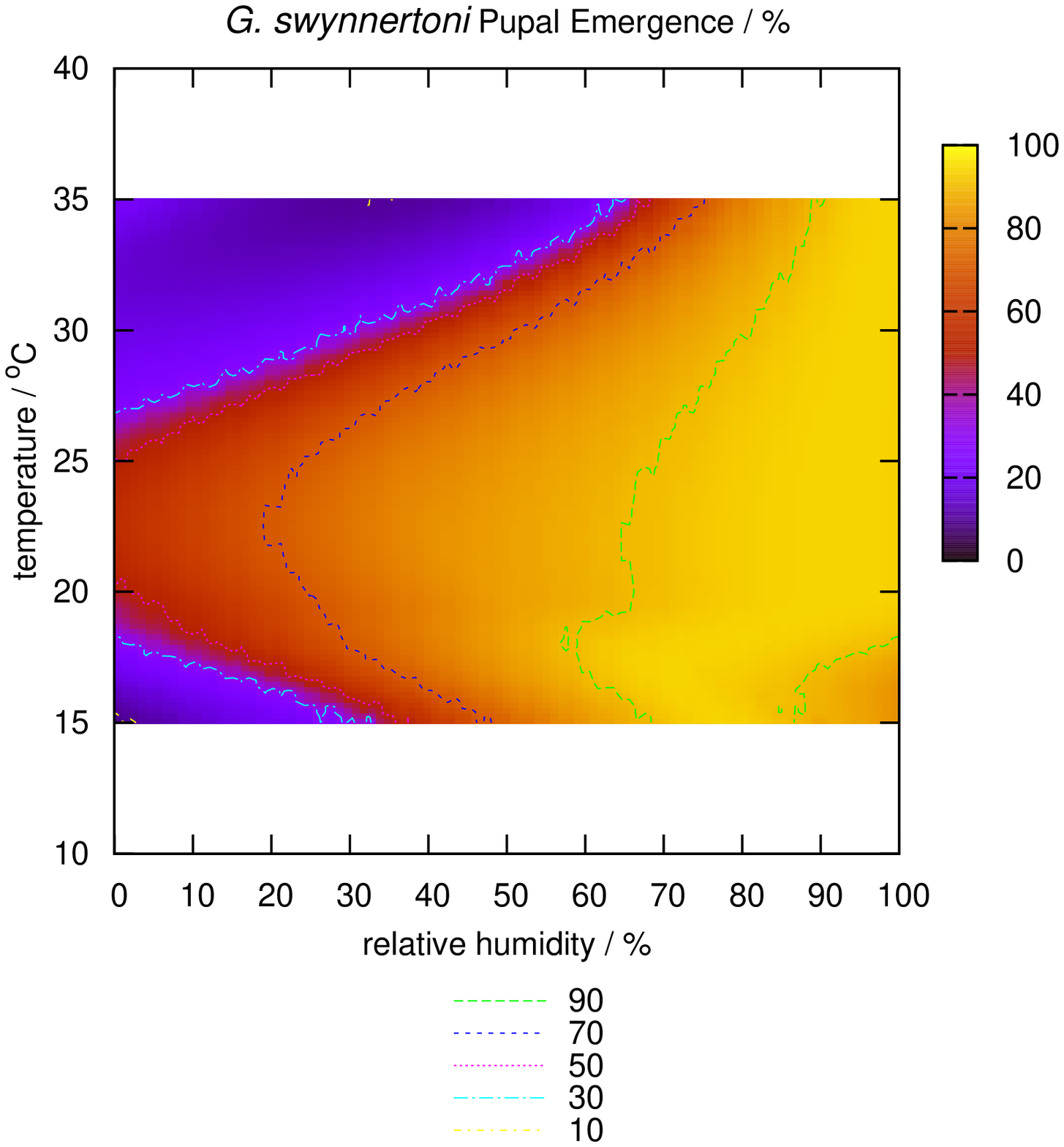}
\caption{{\em G. swynnertoni} pupal emergence (top left) and water loss (top right); {\em G. swynnertoni} pupal emergence for a 35~$^\circ\mathrm{C}$ and 25\% $\mathrm{r.h.}$ heat wave on the first two days after larviposition (bottom left) and on day fifteen and sixteen (bottom right).} \label{swynnertoniHeatWave}
   \end{center}
\end{figure}

Validating the model presents something of a challenge. The deficiency in data,
synonymous with the need for a model, is extreme in this case. Almost all
available data have been incorporated into the model. The veracity of the
results is, to a certain extent, suggested by consistency with the model itself.
For example, transecting the Figs.
\ref{morsitansHeatWave}--\ref{swynnertoniHeatWave} surfaces of emergence at
24~$^\circ\mathrm{C}$ should replicate the Gaussian curves in Fig.
\ref{allSpeciesTogether} and it does. Predicted pupal mortalities due to water
loss should also, logically, never exceed any pupal mortalities observed in the
field for similar humidity and temperature conditions. It is also relevant to
Figs. \ref{austeniHeatWave} and \ref{brevipalpisHeatWave} that Onderstepoort
Veterinary Institute (O.V.I.) keep their \mbox{\em G. austeni} and \mbox{\em G.
brevipalpis} colonies at \mbox{75\% $\mathrm{r.h.}$} (De Beer\nocite{Chantel},
2013).

One set of data on which the model is not based is the measured initial water
reserves for the various species (Table \ref{reserves}). In a world in which
normal distributions are assumed in the absence of any other information, the
measured, critical water loss contour should correspond to that of the 50\%, emergence contour. The sensitivity of {\em G. brevipalpis} and
\mbox{{\em G. palpalis}} pupae to dehydration makes these species arguably the
most challenging tests, as well as of particular interest to this research. {\em
G. brevipalpis} is, furthermore, a topic of intense interest in South 

Africa at the moment and {\em G. palpalis} is known to be a major culprit in the
spread of human trypanosomiasis. Information on the fourth instar excretions is
available for both species, as well as for {\em G. morsitans}. The critical
water reserve is also known for three other species for which, it is hoped, {\em
G. morsitans}-proportionate, fourth instar excretions will suffice. The three
species in question are {\em G. austeni}, {\em G. pallidipes} and {\em Glossina
swynnertoni}.
\begin{table}[H]
\begin{center}
\begin{tabular}{c c c c c c}  
{\em G. austeni} & {\em G. brevipalpis} & {\em G. morsitans} & {\em G. pallidipes} & {\em G. palpalis} & {\em G. swynnertoni} \\
&  &  &  &  & \\ 
5.3 $\mathrm{mg}$ & 18.7 $\mathrm{mg}$ & 8.8 $\mathrm{mg}$ & 10.5 $\mathrm{mg}$ & 7.7 $\mathrm{mg}$ & 8.5 $\mathrm{mg}$ \\ 
\end{tabular}
\end{center}
\caption{Initial water reserves after Bursell\nocite{Bursell1} (1958).} \label{reserves}
\end{table}
A heat wave was also simulated and more especially, the timing thereof, was
experimented with for reasons expounded in the conclusion and which have their
origins in Section \ref{otherSpecies}. A hypothetical heat wave of
35~$^\circ\mathrm{C}$ and 25\% r.h., lasting two days, was based on the kind of
atmospheric temperatures and humidities one might expect in {\em G. morsitans}
habitat. The heat wave was modelled to alternatively coincide with the first two
days after parturition and day fifteen and sixteen of the puparial duration,
these times usually occurring within the unprotected and sensu strictu pupal
stages, respectively. In what way atmospheric conditions relate to those within
the substrate of larviposition is not really known, however, the mesophilic and
xerophilic results may shed some light on the matter.

The same axes for the presentation of the results are used throughout, to
facilitate easy comparison. The reason that results above 35~$^\circ\mathrm{C}$
are blanked out in relevant plots, is that heat wave experiments above
35~$^\circ\mathrm{C}$ are obviously meaningless insofar as the ``heat wave''
still being a heat wave is concerned. Relevant results below
15~$^\circ\mathrm{C}$ are similarly blanked out, since day fifteen and sixteen
no longer occur within the sensu strictu pupal stage once the temperature falls
to a little above 15~$^\circ\mathrm{C}$. Although a hypothetical, late-stage acclimation was experimented with, the results were deemed not significant enough to include.

\section{Conclusions}

This research gives rise to a series of integrals and an algorithm which predict
pupal water loss and consequent survival, given the prevailing soil temperatures
and humidities. High transpiration rates are generally a consequence of high
temperatures and low humidities. They lead to a dehydration of the tsetse pupa
which can be fatal. Although the diametrical opposite is true of transpiration
rates at low temperatures, metabolic processes are slowed, the puparial duration
becomes too long and the cumulative effect of transpiration can be just as
fatal. The low transpiration rates, associated with low temperatures, are more
than compensated for by the lengthening of the duration over which they prevail.
High water losses, in the days immediately following larviposition, also trigger
a more conservative mode of transpiration during the sensu strictu pupal stage
(Bursell\nocite{Bursell1}, 1958, and Childs\nocite{Childs2}\nocite{Childs2a},
2009 and 2009a). 
%

The correspondence between actual, measured critical water losses and those
predicted by the model is such that very little error is discernable (comparing
the position of the critical water loss contour with the 50\%,
emergence contour in Figs. \ref{morsitansHeatWave}--\ref{swynnertoniHeatWave}).
In fact, the results could be said to be remarkable in this regard. It would
therefore appear that the behavioural similarities between different species of
pupae far exceed initial expectations. The agreement between measured and
calculated critical water losses reinforces the assumption that the
water-management strategies and the metabolic time-tables for development differ
little from species to species. This is since the premise on which the {\em G.
morsitans} model is extrapolated to other species, is that the only differences
between species are their relative surface areas, the relative permeability of
their unprotected-stage and protected-stage integuments and their different
fourth instar excretions. One might therefore suspect that all tsetse species
share a common strategy to actively minimise water loss for the majority of
modern habitats and have hydrational mechanisms preventative of dehydration.
Despite the good correspondence in measured and calculated critical water
losses, some caution may still be necessary at temperatures remote from
24.7~$^\circ\mathrm{C}$. The relative permeability of the species' membranes
could change at such extremes.

The {\em G. morsitans}-based model seems to provide a reasonably reliable and
concise definition of hygrophilic species, by way of the factors which
facilitate its extrapolation to other species (in Table
\ref{modelConversionFactors}). The ratio of the unprotected to protected
conversion factors is curiously close to two for all hygrophilic species,
whereas it is around unity for the mesophilic and xerophilic species (it could
suggest the latter categories have a layer of some, or other, protection which
is twice as thick, during the instars). Both exceptions to this rule, \mbox{\em
G. submorsitans} and \mbox{\em G. longipennis}, occur in Sudan as well as
Ethiopia and the aforementioned ratio is remarkably similar, having a value of
around 1.5. A third denizen of these climes, {\em G. tachinoides}, has an
extraordinarily mesophilic pupa for a {\em palpalis} group fly and the
30~$^\circ\mathrm{C}$ exception in the Fig. \ref{allSpeciesTogether} data can be
misleading. Of course, both {\em G. tachinoides} and {\em G. submorsitans} also
have habitat in Chad (Ford and Katondo\nocite{FordAndKatondo}, 1977) and it is
such habitats that elicit interest in a hot, dry spell. 

There would therefore appear to be a certain merit in advancing the course of
this enquiry by way of experimenting with a simulated, two-day heat wave, the
intention being to further elucidate the classification of species as
hygrophilic, mesophilic or xerophilic and, thereby, contribute novel, biological
insight. The xerophilic and mesophilic categories are found not to be as
distinct from one another in such a context as the hygrophilic category is. The
classification of tsetse into separate mesophilic and xerophilic entities could,
possibly, even be considered artificial, in the sense of the one merely being a
more extreme adaption. The responses in \mbox{Figs. \ref{morsitansHeatWave}},
\ref{pallidipesHeatWave} and \ref{swynnertoniHeatWave} are all qualitatively
similar. One clear distinction to emerge between them and the hygrophilic
species is that the latter are far more vulnerable to dehydration during the
early stages not protected by the thin, pupal skin inside the puparium. If the
onset of the heat wave is timed to coincide with the days immediately following
larviposition, instead of during the sensu strictu pupal phase, the effect on
hygrophilic species is profoundly detrimental. In the case of {\em G.
brevipalpis} it is nothing short of catastrophic (Fig.
\ref{brevipalpisHeatWave}). In contrast, the suggestion for mesophilic and
xerophilic species is that such a heat wave is only mildly detrimental and,
even then, mostly effects the higher levels of survival. More than one mechanism
is thought to be responsible for the mesophilic and xerophilic response. One is
the historical conditioning of the transpiration rate (already referred to), a
mechanism by which hot, dry weather tempers the puparium against later losses.
Although sensu strictu pupal transpiration rates are an order of magnitude lower
than the earlier transpiration rates which condition them, any difference in
them prevails for a much longer duration and so the cumulative effect is
potentially as, or more, damaging. Higher transpiration rates associated with a
hot and dry spell are also compensated for, to a certain extent, by a metabolic
quickening of the relevant part of the third and fourth instars over which they
prevail (although, conversely, under certain circumstances, the faster
metabolism allows less time for early acclimation, in this model). For
hygrophilic species, unprotected transpiration rates are simply too high for
either mechanism to make a difference. A hot, dry spell in the days immediately
following larviposition is profoundly detrimental to hygrophilic species. Such
is the strength of this conclusion that further experimentation with late
acclimation suggests the phenomenon to be as irrelevant to hygrophilic species
as the sensu strictu pupal stage is, itself.

A lack of any early conditioning of the puparium gives rise to one further,
surprising feature of mesophilic and xerophilic species: A hot, dry spell is
more detrimental when it coincides with the sensu strictu pupal stage, than when
it occurs immediately after larviposition e.g. \mbox{Fig.
\ref{morsitansHeatWave}}. This certainly is counter-intuitive when the
transpiration rates in Bursell\nocite{Bursell1} (1958) \mbox{Fig. 1} are
contemplated in isolation, as well as being in stark contrast to the hygrophilic
responses (Figs. \ref{austeniHeatWave}--\ref{brevipalpisHeatWave}). Yet, however
compelling it may be to propose an artefact, it would be difficult to refute the
existence of the phenomenon in the case of the {\em G. morsitans} calculation
for dry air at room temperature. Such adaption begs the question of whether an
heat wave could be expected to manifest itself in a different way in the pupal
environment, at different stages of pupal development. Just how the atmospheric
conditions associated with a heat wave manifest themselves in soil, rot holes,
compost etc. is unknown, however, one would certainly expect atmospheric
conditions to prevail at the start of the third instar. Thereafter, one might
wonder about a departure from atmospheric conditions, the pupal environment
having become more insulated. This might explain the phenomenon if, indeed, it
is an adaption. Of course, a more prosaic explanation might be that the
mesophilic and xerophilic species have the capability to survive adversely hot
and dry conditions. They respond by taking countermeasures against any further,
unnecessary losses, preparing themselves for the worst; something which
may not then materialise. For the hygrophilic species, however, the opposite is
true. For the hygrophilic species, things are as expected: A hot, dry spell
which occurs immediately after larviposition is more detrimental than one
coinciding with the sensu strictu pupal stage. 

On these grounds one might argue an apparent qualitatively-different response
for hygrophilic species. For the hygrophilic species it can certainly be said
that third and fourth instar water losses are extremely high. Not only can they
be as much as ten times those of \mbox{\em G. morsitans}, perhaps more important
is the fact that the ratio of the unprotected to protected conversion factors
(in Table \ref{modelConversionFactors}) is double the same ratio for the
mesophilic and xerophilic species. At what point these early losses render the
long duration of the sensu strictu pupal stage irrelevant is not clear. Whether
a milder heat wave would reproduce the mesophilic and xerophilic response in
hygrophilic species, is not known. Whether a milder heat wave would induce the
same preference for an early, rather than later, exposure to hot and dry
conditions is a question not answered in this research. One would not expect
atmospheric conditions anywhere near as dry as those used for the hypothetical
heat wave in {\em G. austeni} habitat, although they certainly do occur in some
\mbox{\em G. brevipalpis} country. Despite this, the soil humidity and soil
temperature within the levees and river terraces of large drainage lines, which
are so often the habitat of {\em G. brevipalpis}, could differ altogether from
those which characterise the atmosphere. For that matter, it is not known how
atmospheric conditions manifest themselves in the pupal environments of any of
the other species, either. In many tsetse habitats the mean daily temperatures
seldom change by more than two degrees at a time, and a change of four degrees
from one day to the next may be considered extreme. The transition to and from
the heat wave condition is therefore completely unrealistic for large parts of
the domain investigated. Experimenting in such a manner does, however, allow the
conclusions to be generally stated and so makes for an interesting study
nonetheless. One might otherwise have been tempted to conclude that there is a
continuous progression from {\em G. pallidipes} to {\em G. austeni}, that {\em
G. austeni} is simply a more extreme adaption. It is not, if the experiments
with the hypothetical heat wave can be regarded as relevant. If one considers
this pupal work, in conjunction with the work pertaining to the teneral stage
(Childs\nocite{Childs6}, 2014), it points to one inevitable conclusion: That the
classification of species as hygrophilic, mesophilic and xerophilic is largely
true only in so much as their third and fourth instars are and, possibly, the
hours shortly before eclosion. The fate of hygrophilic species is largely
decided by the conditions which prevail during the third and fourth instars and,
possibly, the hours shortly before eclosion.

An explanation for the `squiggle' associated with low temperature in the lower,
right plots of Figs. \ref{morsitansHeatWave}--\ref{swynnertoniHeatWave} might be
in order. At very low temperatures the puparial duration becomes so long that an
heat wave, generally timed to occur during the sensu strictu pupal phase, falls
within the fourth instar, instead. As the temperature drops and the metabolism
slows, the puparial duration lengthens. The steep transition zone between the
unprotected and protected transpiration rates (evident in Fig.
\ref{mixOfRatesWithTime}) begins to enter the fifteenth day, the day scheduled
for the commencement of the intended, sensu strictu pupal-stage heat wave. The
pupa then begins to incur the massive water losses that a lack of waterproofing
implies. At some point this rapid transition in rates enters the heat wave just
enough for the optimum water loss to be incurred. At some point the water loss
is just sufficient to produce a maximum conditioning for a minimum tax on
reserves, thereafter it becomes more damaging; hence, the `squiggle' in the
results which occurs between 15~$^\circ\mathrm{C}$ and 20~$^\circ\mathrm{C}$. At
a little above 15~$^\circ\mathrm{C}$, the coincidence of the fourth instar and
the heat wave becomes complete. The heat wave no longer coincides with any part
of the sensu strictu pupal stage, as intended, hence the omission of the results
at temperatures any lower than 15$^\circ$C. 

The real value of the Terblanche and Kleynhans\nocite{TerblancheAndKleynhans1}
(2009) experiment is that it suggests acclimation is still possible well after
the onset of the sensu strictu pupal stage, as well as reinforcing the
assumption that the transpiration rate of many species is roughly a multiple of
the {\em G. morsitans} rate. A late-acclimation effect, along the lines of the
effect in Bursell\nocite{Bursell1} (1958), was crudely implemented for the sensu
strictu pupal stage. Experimenting with this added level of complexity revealed
no obviously discernable differences in the results for hygrophylic species. For
{\em G. morsitans}, the high-temperature boundary for survival (the 50\% and
30\% emergence contours) moved to a position 3~$^\circ\mathrm{C}$ higher, into
the minor, shallow and flat band, across the hot and dry corner of Fig.
\ref{morsitansHeatWave}. The reason for the slight upward shift in the relevant
contours is thought to be that early acclimation can now be completed in the
sensu strictu pupal stage. The only other difference caused by a late
acclimation, was that the xerophilic and mesophilic preference for the timing of
a heat wave during the third and fourth instars, became less apparent.  

The model formulated is obviously intended for more ambitious purposes than the
mere interpretation and visualization of data. It, nonetheless, proves to be
an invaluable tool in the interpretation and visualization of the
Bursell\nocite{Bursell1} (1958) endeavour. A substantial body of literature is
of the opinion that pupal mortality due to dehydration is either irrelevant, can
be assumed constant, is linearly dependent on temperature, or is dependent on
temperature alone. Dehydration does tend to be more temperature-dependent in the
mesophilic and xerophilic species (e.g. Fig. \ref{morsitansHeatWave}), although,
even then, that dependence is certainly not linear. Notice that even in the {\em
morsitans} group, daily pupal mortality is neither linear, nor a function of
temperature alone. Even for a hardy fly such as {\em G. morsitans}, its
prospects deteriorate rapidly once out of favourable habitat. When it comes to a
species such as {\em G. brevipalpis}, however, there would be more merit in
assuming dehydration-related survival to be  entirely humidity-dependent
(\mbox{Fig. \ref{brevipalpisHeatWave}}). When it comes to hygrophilic species,
that pupal mortality due to dehydration is both relevant and palpable is beyond
contention. The effect of soil humidity is profound. Soil humidity defines
habitat. This is despite the fact that water loss and any consequent pupal
mortality are also very different things (water loss may culminate in and
ultimately take its toll on the teneral). If it could be said that there was a
single, overriding fact that the Bursell\nocite{Bursell1} (1958) investigation
was able to reveal, it is that the water reserve is a limiting factor in tsetse
pupae. Dehydration is a challenge to pupae, if not a major threat and it is
generally accepted that most of the {\em Glossina} genus is not well adapted to
arid environments (Glasgow\nocite{Glasgow1}, 1963). The {\em Glossina} genus may
well derive from a common, tropical, rain-forest dwelling ancestor, adjusted to
moist, warm climates (Glasgow\nocite{Glasgow1}, 1963). 

Notice that this is not to claim that pupal mortality due to dehydration is
always decidedly higher than any other causes of mortality, only that causes of
higher mortality are likely to be geographically more uniform, random, or
cyclical. There may also be additional mortality, only indirectly attributable
to dehydration. A dearth of humid substrates might lead to more localised and
concentrated larviposition, consequently, to increased predation and parasitism.
The results in Figs. \ref{morsitansHeatWave}--\ref{swynnertoniHeatWave} point to
the fact that pupal sites for some species are very much confined in the dry
season, particularly in the case of South Africa's two extant species, {\em G.
austeni} and \mbox{\em G. brevipalpis}. These would be obvious places in which
to concentrate control measures and one immediate application of this research.
Barriers of the type modelled in Childs\nocite{Childs3} (2010) might be far more
efficacious if deployed in the immediate vicinity of pupal sites, rather than
for the purposes of containment. While early stage mortality is considered to be
the most significant, by far, in any model of tsetse population dynamics it is
of even greater relevance when in the context of control measures. This is since
the pupal stage, alone, is neither susceptible to targets, nor aerial spraying.
In this regard, it is noteworthy that the eclosion rates in
Childs\nocite{Childs4} (2011) should probably be closer to those in
Childs\nocite{Childs5} (2013), if not higher, for a steady-state equilibrium and
the projected outcomes can be adjusted by a very similar factor. The humidities
and temperatures referred to in this research have generally been attributed to
the pupal substrate, however, one might still wonder whether atmospheric
conditions can be wholly ignored. Hygrophilic species' apparent vulnerability
during the third instar, coupled with the discovery that atmospheric conditions
may sometimes be profoundly different from those which characterise the pupal
substrate (Childs\nocite{Childs6}, 2014) are cause for concern, although some
\mbox{\em G. brevipalpis} environments suggest the opposite. The habitats of
certain tsetse species might otherwise be used in conjunction with Figs.
\ref{morsitansHeatWave}--\ref{swynnertoniHeatWave} as an indicator of soil
moisture content. Some caution still needs to be exercised, as
Childs\nocite{Childs6} (2014) reveals why the highly eccentric strategy of {\em
G. brevipalpis} (\mbox{Fig. \ref{brevipalpisHeatWave}}) succeeds, as well as how
the equally contradictory pupal and adult strategies of {\em G. longipennis}
perfectly complement those of \mbox{\em G. brevipalpis}. The pupal and adult
strategies of \mbox{\em G. austeni} complement those of \mbox{\em G.
brevipalpis} to a lesser extent, although sufficiently well to allow their
frequent sympatric association.

The results in Figs. \ref{morsitansHeatWave}--\ref{swynnertoniHeatWave} must
ultimately be described as projections, nonetheless. They are no substitute for
good, field data, were such data to exist. The historical conditioning of the
puparium, in all species, has been based proportionately on that of {\em G.
morsitans}, as has a \mbox{\em G. morsitans}-proportionate excretion been used
for the other species belonging to the {\em morsitans} group. It would now
appear that the modelled acclimation is simplistic, even for {\em G. morsitans}.
Acclimation would, however, appear to be irrelevant in the case of hygrophilic
species. Consider then, for instance, that if Bursell\nocite{Bursell1} (1958) is
mistaken insofar as it attributes waterproofing to being the cause of the
precipitous drop in transpiration rates concomitant with the end of the fourth
instar, if that drop was, instead, actually the result of a puparium, of finite
water content, drying and the timing thereof coincidental, such a mistake might
introduce large errors to the calculation. Such pessimism should be tempered by
the observation that the model has already been demonstrated sufficiently sound,
to the extent that any contrary data can be used to effect improvements. For
example, any data which would demonstrate \mbox{Assumption \ref{assumption6}} to
be problematic could be used to replace $\frac{dt}{d {\bar t}}(T)$ with a new
function, $\frac{dt}{d {\bar t}}({\bar t}, T)$, predicting how the different
parts of the puparial duration change with temperature. The same data which
would demonstrate Assumption \ref{assumption3} to be problematic at the extrema
of the temperature domain, could similarly be used to replace $\delta$ with a
new relative-to-{\em G. morsitans}-rate, $\delta(T)$, one dependent on
temperature. Although the assumptions are simplistic, the model, at very least,
elucidates the trends and the suggestion is that simplicity is all that is
needed for hygrophilic species. The need for a better understanding of
acclimation and related experimental data is possibly also of importance.
Lastly, Phelps\nocite{Phelps1} (1973) substantiated the Bursell\nocite{Bursell1}
(1958) claim that the laboratory pupae in question were slightly inferior,
showing that they only correspond to pupae in the wild during the most
unfavourable season. Consider, however, that if a model is able to be adapted
and successfully make predictions with respect to other species, how much more
suitable must it be for adaption to the phenotypic plasticity within the same
species. Since pupal dehydration would appear to be the most challenging aspect
of modelling early stage mortality, the prognosis for this model would be one of
greater significance than any problems arising from issues such as inferior
quality pupae and differing puparial durations and the shortage of statistically
significant data can be corrected at some stage. It is with the remainder of the
pupal reserves that the newly-eclosed teneral fly either hops onto a hock, for
example, at sunset (Vale et. al.\nocite{ValeHargroveJordanLangleyAndMews} 1976)
or more likely, waits through the night until dawn to feed; the topic of
Childs\nocite{Childs6} (2014). 

\section{Acknowledgements} 

The author is indebted to Neil Muller, as well as to Schalk Schoombie, Johan Meyer and Glen Taylor for hosting this research.
\bibliography{improvedPupalH2Oloss}

\begin{thebibliography}{10}

\bibitem{Bursell1}
E.~Bursell.
\newblock The water balance of tsetse pupae.
\newblock {\em Philosophical Transactions of the Royal Society of London},
  241(B):179--210, 1958.

\bibitem{Bursell2}
E.~Bursell.
\newblock The water balance of tsetse flies.
\newblock {\em Transactions of the Royal Entomological Society London},
  111:205--235, 1959.

\bibitem{Bursell3}
E.~Bursell.
\newblock The effect of temperature on the consumption of fat during pupal
  development in {{\em {G}}}{\em lossina}.
\newblock {\em Bulletin of Entomological Research}, 51(3):583--598, 1960.

\bibitem{BuxtonAndLewis1}
P.~A. Buxton and D.~J. Lewis.
\newblock Climate and tsetse flies: laboratory studies upon {{\em {G}lossina
  submorsitans}} and {{\em tachinoides}}.
\newblock {\em Philosophical Transactions of the Royal Society of London},
  224(B):175--240, 1934.

\bibitem{Childs2a}
S.~J. Childs.
\newblock A simple model of pupal water loss in {{\em {G}lossina}}.
\newblock {\em arXiv preprint http://arxiv.org/abs/0901.2470}, \ 2009.

\bibitem{Childs2}
S.~J. Childs.
\newblock A model of pupal water loss in {{\em {G}lossina}}.
\newblock {\em Mathematical Biosciences}, 221:77--90, 2009.

\bibitem{Childs3}
S.~J. Childs.
\newblock The finite element implementation of a {K.P.P.} equation for the
  simulation of tsetse control measures in the vicinity of a game reserve.
\newblock {\em Mathematical Biosciences}, 227:29--43, 2010.

\bibitem{Childs4}
S.~J. Childs.
\newblock Theoretical levels of control as a function of mean temperature and
  spray efficacy in the aerial spraying of tsetse fly.
\newblock {\em Acta Tropica}, 117:171--182, 2011.

\bibitem{Childs5}
S.~J. Childs.
\newblock A set of discrete formulae for the performance of a tsetse population
  during aerial spraying.
\newblock {\em Acta Tropica}, 125:202--213, 2013.

\bibitem{Childs6}
S.~J. Childs.
\newblock A model of teneral dehydration in {{\em {G}lossina}}.
\newblock {\em Acta Tropica}, 131:79--91, 2014.

\bibitem{Chantel}
C.~De~Beer.
\newblock {\em By communication}.
\newblock 2013.

\bibitem{DuToit}
R.~Du~Toit.
\newblock Trypanosomiasis in {Z}ululand and the control of tsetse flies by
  chemical means.
\newblock {\em Onderstepoort Journal of Veterinary Research}, 26(3):317--387,
  1954.

\bibitem{FordAndKatondo}
J.~Ford and K.~M. Katondo.
\newblock The distribution of tsetse flies in {A}frica (3 maps).
\newblock {\em OAU, Cook, Hammond \& Kell, Nairobi}, 1977.

\bibitem{Glasgow1}
J.~P. Glasgow.
\newblock {\em The Distribution and Abundance of Tsetse}.
\newblock International Series of Monographs on Pure and Applied Biology.
  Pergamon Press, 1963.

\bibitem{Hargrove1}
J.~W. Hargrove.
\newblock Age--dependent changes in the probabilities of survival and capture
  of the tsetse, {{\em Glossina {M}orsitans {M}orsitans {W}estwood}}.
\newblock {\em Insect Science and its Applications}, 11(3):323--330, 1990.

\bibitem{Hargrove3}
J.~W. Hargrove.
\newblock {\em The Trypanosomiases}.
\newblock Editors: I. Maudlin, P. H. Holmes and P. H. Miles. {CABI} publishing,
  {O}xford, U.K., 2004.

\bibitem{Parker1}
A.~Parker.
\newblock {\em By communication}.
\newblock 2008.

\bibitem{Phelps1}
R.~J. Phelps.
\newblock The effect of temperature on fat consumption during the puparial
  stages of {{\em {G}lossina morsitans morsitans}} westw. (dipt. glossinidae)
  under laboratory conditions, and its implication in the field.
\newblock {\em Bulletin of Entomological Research}, 62:423--438, 1973.

\bibitem{phelpsAndBurrows1}
R.~J. Phelps and P.~M. Burrows.
\newblock Puparial duration in {{\em {G}lossina morsitans orientalis}} under
  conditions of constant temperature.
\newblock {\em Entomologia Experimentalis et Applicata}, 12:33--43, 1969.

\bibitem{RogersAndRobinson}
D.~J. Rogers and T.~P. Robinson.
\newblock {\em The Trypanosomiases}.
\newblock Editors: I. Maudlin, P. H. Holmes and P. H. Miles. {CABI} publishing,
  {O}xford, U.K., 2004.

\bibitem{RogersAndRandolph1}
David~J. Rogers and Sarah~E. Randolph.
\newblock Estimation of rates of predation on tsetse.
\newblock {\em Medical and Veterinary Entomology}, 4:195--204, 1990.

\bibitem{TerblancheAndKleynhans1}
J.~S. Terblanche and E.~Kleynhans.
\newblock Phenotypic plasticity of dessication resistance in {{\em {G}}}{\em
  lossina} puparia: Are there ecotype constraints on acclimation responses?
\newblock {\em Journal of Evolutionary Biology}, 22(8):1636--1648, 2009.

\bibitem{ValeHargroveJordanLangleyAndMews}
G.~A. Vale, J.~W. Hargrove, A.~M. Jordan, P.A. Langley, and A.~R. Mews.
\newblock Survival and behaviour of tsetse flies ({{\em {D}iptera}}, {{\em
  {G}lossinidae}}) released in the field: a comparison between wild flies and
  animal-fed and in {\em vitro}-fed laboratory-reared flies.
\newblock {\em Bulletin of Entomological Research}, 66(4):731--744, 1976.

\end{thebibliography}

\end{document}